\documentclass[aps,pra,preprint,superscriptaddress,longbibliography]{revtex4-2}
\usepackage{graphicx}
\usepackage{amsfonts}
\usepackage{amsmath}
\usepackage[usenames,dvipsnames]{color}
\usepackage{bm}
\usepackage{amssymb}

\usepackage{multirow}
\usepackage{epstopdf}
\usepackage{natbib}

\usepackage{hyperref}
\hypersetup{
colorlinks=true,
linkcolor=blue,
citecolor=blue,
filecolor=green,
urlcolor=blue,
}

\newcommand{\bra}[1]{\langle #1 |}
\newcommand{\ket}[1]{| #1 \rangle}

\newcommand{\beq}{\begin{equation}}
\newcommand{\eeq}{\end{equation}}

\usepackage{algorithm}
\usepackage{algpseudocode}

\begin{document}

\title{Total angular momentum representation for state-to-state quantum scattering of cold molecules in a magnetic field}

\author{Suyesh Koyu}
\affiliation{Department of Physics, University of Nevada, Reno, NV, 89557, USA}
\author{Rebekah Hermsmeier}
\affiliation{Department of Physics, University of Nevada, Reno, NV, 89557, USA}
\author{Timur V. Tscherbul}
\affiliation{Department of Physics, University of Nevada, Reno, NV, 89557, USA}

\date{\today}

\begin{abstract}
We show that the integral cross sections for state-to-state quantum scattering of cold molecules in a magnetic field can be efficiently computed using the total angular momentum representation despite the presence of unphysical Zeeman states in the eigenspectrum of the asymptotic Hamiltonian. We demonstrate that the unphysical states arise due to the incompleteness of the space-fixed total angular momentum basis caused by using a fixed cutoff value $J_\text{max}$ for the total angular momentum of the collision complex $J$. As a result, certain orbital angular momentum ($l$) basis states lack the full range of $J$ values required by the angular momentum addition rules, resulting in the appearance of unphysical states. We find that by augmenting the basis with a full range of $J$-states for every $l$, it is possible to completely eliminate the unphysical states from quantum scattering calculations on molecular collisions in external fields. To illustrate the procedure, we use the augmented basis sets to calculate the state-to-state cross sections for rotational and spin relaxation in cold collisions of $^{40}$CaH$(X^2\Sigma^+,v=0,N=1, M_N=1, M_S=1/2)$ molecules with $^4$He atoms in a magnetic field. We find excellent agreement with benchmark calculations, validating our proposed procedure. We find that $N$-conserving spin relaxation to the lowest-energy Zeeman state of the $N=1$ manifold, $|11\frac{1}{2}\rangle\to |1-1-\frac{1}{2}\rangle$ is nearly completely suppressed due to the lack of spin-rotation coupling between the fully spin-stretched Zeeman states. Our results demonstrate the possibility of rigorous, computationally efficient, and unphysical state-free quantum scattering calculations on cold molecular collisions in an external magnetic field.

\end{abstract}

\maketitle

\newpage

\section{Introduction}

Ultracold molecular collisions offer a rich playground for exploring and controlling fundamental quantum phenomena in ultracold dilute molecular gases \cite{Bohn:17}. These fascinating phenomena range from resonant tunneling \cite{Balakrishnan:01}, external field-induced reactive scattering resonances \cite{Tscherbul:15,Hermsmeier:21} and long-lived reaction complex formation \cite{Croft:14,Liu:20}  to quantum coherent control \cite{Devolder:21}, many-body entanglement \cite{Bilitewski:21}, and quantum chaos \cite{Croft:17}. In addition, the strong anisotropic dipole-dipole interactions of cold polar molecules are highly tunable by external  electric and microwave fields, enabling applications in quantum information processing \cite{DeMille:02,Andre:06,Yelin:06}, quantum metrology \cite{Bilitewski:21} and quantum simulation \cite{Gorshkov:11,Hazzard:13}.




First-principles quantum dynamics calculations based on {\it ab initio} potential energy surfaces (PESs) \cite{Volpi:02,Krems:04,Tscherbul:09,Tscherbul:09c,Tscherbul:10,Tscherbul:11,Suleimanov:12,Campbell:09,Hummon:11,Bergeat:15,Vogels:15,Bergeat:18,Klein:17,Vogels:18,Morita:18,Morita:20} provide the ultimate theoretical understanding of cold and ultracold molecular collisions allowed by the laws of quantum mechanics.  By providing the full state-to-state information about collision observables in the presence of external fields, such calculations can be used to explain experimentally observed phenomena in ultracold molecular gases \cite{Campbell:09,Hummon:11,Bergeat:15,Vogels:15,Bergeat:18,Klein:17,Vogels:18} and to predict novel effects in molecular systems not yet realized experimentally \cite{Lara:06,Wallis:09,Pavlovic:09,Tscherbul:09,Tscherbul:09c,Zuchowski:11,Tscherbul:06,Tscherbul:11,Morita:17,Morita:18,Morita:20}. Quantum dynamics calculations could be based on either time-independent or time-dependent scattering formalisms. In the time-dependent approach that has been most widely adopted thus far \cite{Volpi:02,Krems:04,Tscherbul:09,Tscherbul:09c,Tscherbul:10,Tscherbul:11,Suleimanov:12,Campbell:09,Hummon:11,Bergeat:15,Vogels:15,Bergeat:18,Klein:17,Vogels:18,Morita:18,Morita:20} the wavefunction of the collision complex is expanded in a channel basis set \cite{ColdChem:18}, and the resulting coupled-channel (CC) equations are solved numerically to generate the full state-to-state scattering $S$-matrix, which provides  comprehensive information about the scattering properties (such as the differential/integral cross sections and collision rates) of molecules in different internal states. 

The choice of the channel basis is a central aspect of any quantum dynamics calculation, be it time-dependent or time-independent.  An efficient basis set results in a compact representation of the scattering wavefunction using the fewest possible basis functions.
One such basis set, which proved particularly efficient for molecular collisions in external fields, is based on the total angular momentum representation in either the body-fixed (BF) or space-fixed (SF) coordinate frames \cite{Tscherbul:10,Tscherbul:12,Suleimanov:12}.



Even though the quantum theory of molecular collisions in external fields was first formulated in the uncoupled SF representation \cite{Volpi:02,Krems:04}, it was soon realized that this representation is not computationally efficient for strongly anisotropic atom-molecule and molecule-molecule collisions, where the interaction anisotropy couples an exceedingly large number of basis functions, making converged calculations impossible.  In contrast, the total angular momentum representation introduced by Arthurs and Dalgarno in 1960 for field-free collisions \cite{Arthurs:60} and by our group in 2010-2012 for collisions in external fields \cite{Tscherbul:10,Suleimanov:12,Tscherbul:12}, takes into account the approximate conservation of the total angular momentum of the collision complex to drastically reduce the number of basis states required to obtain converged scattering observables.  The total angular momentum basis has since been used to obtain converged results for a variety of ultracold atom-molecule \cite{Tscherbul:11,Morita:18,Morita:20}, molecule-molecule \cite{Suleimanov:12}, and chemically reactive  \cite{Tscherbul:15} systems, which had been considered intractable prior to the development of the basis.

A distinctive feature of the total angular momentum basis is the appearance of  unphysical Zeeman states  of the asymptotic Hamiltonian.
 As shown in our previous work \cite{Tscherbul:10} the unphysical  states correspond to the maximum value of the total angular momentum $J_\text{max}$ included in the basis set. As a result, transitions involving such states are strongly suppressed at ultralow temperatures, so they can be simply ignored in calculating ultracold collision observables, as was done in Refs.~\cite{Morita:17,Morita:18,Morita:20}.
  Nevertheless, the unphysical states 
   tend to congest  the  spectrum of molecular threshold energies, particularly in the vicinity of nearly degenerate manifolds of excited rotational and/or hyperfine states.  
This congestion complicates the identification of molecular thresholds and makes it challenging to calculate state-to-state scattering observables in the presence of external fields. It could also cause problems in calculating the near-threshold bound states of collision complexes in the presence of external fields, an essential step in interpreting  Feshbach resonances in ultracold atom-molecule \cite{Tscherbul:07,Hummon:11,Wang:21} and molecule-molecule \cite{Tscherbul:09c,Suleimanov:11} collisions. Indeed, it is impossible to  distinguish between the unphysical and physical states without knowing the exact energies of the latter (used as a benchmark). Thus, while  identifying unphysical states in diatomic molecules is a relatively straightforward task, it is far from trivial for the near-threshold bound states of atom-molecule and molecule-molecule complexes, motivating an in-depth study into the origin of unphysical states, and the development of efficient strategies for their elimination.


  
  Here, we present such a study using a space-fixed (SF) total angular momentum representation of the scattering Hamiltonian in the presence of an external magnetic field. This representation clarifies the symmetry of the asymptotic Hamiltonian with respect to the orbital motion of the atom around the diatomic molecule, as a result of which the Hamiltonian  commutes with $\hat{{L}}^2$, the squared orbital angular momentum operator of the collision complex. This results in a representation, which block-diagonalizes the matrix of the asymptotic Hamiltonian, and clarifies the origin of the unphysical states as due to the incompleteness of the SF total angular momentum basis.
   This allows us to pinpoint the basis states responsible for the formation of unphysical states, and to suggest a procedure for completing the SF basis, leading to a general strategy for eliminating  the unphysical states from quantum scattering calculations on molecular collisions in magnetic fields.
  
To illustrate the basis set completion procedure, we calculate the collision energy dependence of state-to-state scattering cross sections for cold collisions of rotationally excited CaH molecules with $^4$He atoms in a magnetic field using the augmented SF basis. Our results are   in excellent agreement  with benchmark calculations based on the fully uncoupled SF representation \cite{Volpi:02,Krems:04} for all final states over a wide region of collision energies, demonstrating the feasibility of unphysical state-free quantum scattering calculations with a negligible computational overhead for low-lying rotational states.

This article is organized as follows. Section IIA presents the necessary background information on the total angular momentum representation for molecular collisions in a magnetic field. 
In Sec. IIB we consider the origin of unphysical states in the SF total angular momentum representation, and describe a basis set  completion procedure to arrive at an unphysical-state-free, augmented SF representation of the asymptotic Hamiltonian.  Section IIIA illustrates the procedure by constructing the augmented SF basis for a $^2\Sigma$ molecule in a magnetic field. Quantum scattering calculations of state-to-state cross sections for cold $^{4}$He~+~$^{40}$CaH collisions in the augmented SF basis are presented and compared with benchmark results  in Sec.~IIIB. Finally, Sec. IV summarizes our main findings and outlines a few promising directions of future research.

\section{Theory}

\subsection{Total angular momentum basis for collisions in an external magnetic field}

Here, we will only give  the essential details of the total angular momentum representation for molecular collisions in a magnetic field, which are required to introduce the concept of unphysical states. 
The reader is referred to Refs.~\cite{Tscherbul:10,Suleimanov:12} for a detailed description of the theory and computational methodology. 

In the remainder of this paper, we will focus on the case of cold collisions of $^2\Sigma$ molecules with spherically symmetric atoms in the $^1$S electronic state. Owing to its simplicity, this case is ideally suited to illustrate the main idea of our approach, 
which can be extended to  molecule-molecule collisions and chemical reactions in a magnetic field following \cite{Tscherbul:10,Tscherbul:15,Suleimanov:12}.

 
 Our goal is to solve the time-independent Schr\"odinger equation for the collision complex  formed by a structureless atom and a $^2\Sigma$ molecule and described by the Hamiltonian 
\begin{equation}
\hat{{H}} = - \frac{1}{2\mu R}  \frac{d^2}{dR^2} R
                     + \frac{\hat{{L}}^2}{2\mu R^2} 
                     + \hat{{H}}_{\mathrm{mol}} + \hat{V}(R,\theta),
\label{eq:Heff}
\end{equation}
subject to the scattering boundary conditions. Here, $R$ is the magnitude of the Jacobi vector $\mathbf{R}$ from the center of mass of the diatomic molecule to the atom, $\theta$ is the Jacobi angle between $\mathbf{R}$ and $\mathbf{r}$, the internuclear separation vector of the molecule, $\mu$ is the reduced mass and $\hat{{L}}$ is the orbital angular momentum of the atom-molecule system. Furthermore, $\hat{V}$ is the atom-molecule interaction potential, and $\hat{H}_\mathrm{mol}$ is the Hamiltonian that describes the $^2 \Sigma$ molecule’s internal structure, which includes the rotational structure, the spin-rotation interaction,  and the Zeeman effect 
\begin{equation}
\hat{{H}}_\mathrm{mol} = B_e \hat{N}^2  +\gamma_\text{sr} \hat{\mathbf{N}}\cdot \hat{\mathbf{S}} + 2\mu_0 B\hat{S}_Z,
\label{eq:Hmol}
\end{equation}
where $\hat{\mathbf{N}}$ is the rotational angular momentum of the diatomic molecule, $\hat{\mathbf{S}}$ is the electronic spin, $B_e$ is the rotational constant, $\gamma_\text{sr}$ is the spin-rotation constant, $\mu_0$ is the Bohr magneton, and $B$ is the magnitude of the external magnetic field,  which defines the SF quantization axis.  
We  neglect the hyperfine structure, which is a good approximation in the large-field limit \cite{Morita:18}, and assume that the internuclear distance of the diatomic molecule is fixed at its equilibrium value (the rigid rotor approximation).

To solve the time-independent Schr\"odinger equation with the Hamiltonian (\ref{eq:Heff}) we expand the wavefunction in the basis set \cite{Tscherbul:10}  
\begin{equation}
\label{eq:totjbasis}
|\Psi\rangle = \frac{1}{R} \sum^{}_{n}F_{n}(R) \ket{\Phi_n}, 
\end{equation}
where $ \ket{\Phi_n}$ are the angular basis functions and $F^{M}_{n}(R)$ are the radial expansion coefficients. 




We choose the angular basis functions to be the eigenstates of $|\hat{\mathbf{J}}|^2$ and $\hat{J}_Z$, where $\hat{\mathbf{J}}$ is the total angular momentum of the collision complex, and $\hat{J}_Z$ is its  projection on the SF quantization axis defined by the external (magnetic) field. 
As shown in our previous work,  the SF total angular momentum basis has excellent convergence properties, 
 allowing rigorous numerical treatment of  strongly anisotropic molecular collisions \cite{Tscherbul:10,Tscherbul:11,Suleimanov:12} and chemical reactions \cite{Tscherbul:15} in  the presence of external dc electric and/or magnetic  fields.

 To construct the SF total angular momentum basis functions, we first couple  $\mathbf{N}$ and $\mathbf{S}$ form the total angular momentum  of the isolated $^2\Sigma$ molecule $\hat{\mathbf{j}}$, and then couple $\hat{\mathbf{j}}$ with $\hat{\mathbf{L}}$ to form $\hat{\mathbf{J}}$
\begin{equation}
\label{eq:totjbasisSF}
\ket{\Phi_n} = |JMlj(NS)\rangle = \sum_{m_j,m_l} 
 \langle jm_j, lm_l | JM\rangle
|jm_j(NS)\rangle |lm_l\rangle
\end{equation}
where
where $ |jm_j(NS)\rangle = \sum_{N,M_N} \langle NM_N SM_S|jm_j\rangle |NM_N\rangle |SM_S\rangle $ are the eigenstates of $|\hat{\mathbf{j}}|^2$ and $\hat{j}_Z$, $\ket{NM_N}$ are those of $|\hat{\mathbf{N}}|^2$ and $\hat{N}_Z$, and $\ket{SM_S}$ are those of $|\hat{\mathbf{S}}|^2$ and $\hat{S}_Z$ (for $^2\Sigma$ molecules considered here, $S=1/2$), and $\langle j_1m_1j_2 m_2|jm\rangle$ are the Clebsch-Gordan coefficients.



The radial expansion coefficients $F_n(R)$ in Eq. (\ref{eq:totjbasis}) can be obtained by solving the CC equations 
\begin{equation}\label{CCeqs}
 \Big[ \frac{d^2}{dR^2}+2\mu E \Big] F_{n}(R)=2\mu
 \sum_{n'} \langle n | \hat{V} (R,\theta)+ \frac{\hat{{L}}^2}{2\mu R^2}+ \hat{H}_\text{mol}   | n'\rangle  
  F_{n'}(R)
\end{equation}
where $|n\rangle$ denote the coupled basis states $|JMlj(NS)\rangle$. The asymptotic behavior of the solutions in the limit $R\to \infty$ defines the scattering $S$-matrix \cite{Tscherbul:10} for a given value of the  total angular momentum projection onto the field axis $M$, from which the integral cross sections  are evaluated as
\begin{equation}\label{StoCS}\begin{aligned}
\sigma_{\gamma \to \gamma' }=\frac{2\pi}{k_\gamma^2} \sum_M \sum_{l,l'} |\delta_{\gamma\gamma'}\delta_{ll'} -S^M_{\gamma l, \gamma'l'}|^2,
\end{aligned}
\end{equation}
 where $\gamma$ and $\gamma'$ are the internal states of the colliding molecule (or eigenstates of $\hat{H}_\text{mol}$), among which collision-induced transitions  occur.
 
The CC equations are parametrized by the matrix elements of the molecular Hamiltonian $\hat{H}_\text{mol}$, the interaction potential  $\hat{V}(R,\theta)$, and the angular kinetic energy $\hat{L}^2/(2\mu R^2)$. Below we present the expressions for these matrix elements in the  SF total angular momentum basis, derived using standard techniques of angular momentum theory \cite{Zare:88}

We start from the Zeeman matrix elements
\begin{multline}\label{ZeemanMatrix}
\langle JMlj(NS)|2 \mu_0 B \hat{S}_Z| J’M’ l’j’(N’S’)\rangle=\delta_{ll’} \delta_{N N’} 
(-1)^{J-M+J’+l+2j+N+S+2}
[(2J+1)(2J’+1) \\ \times (2j’+1)(2j+1){(2S+1)S(S+1)}]^{1/2}2\mu_0 B
\begin{pmatrix}
J&1&J’\\
-M & 0 &M
\end{pmatrix}
\begin{Bmatrix}
j&j’&1\\
J’ & J &l
\end{Bmatrix}
\begin{Bmatrix}
S&S&1\\
j’ & j &N
\end{Bmatrix},
\end{multline}
where the symbols in parentheses and figure brackets are the 3-$j$ and 6-$j$ symbols. Note that the interaction with external magnetic fields couples the states with different total angular momenta $J$, a consequence of the breaking of rotational invariance in the presence of external fields.
The matrix elements of the rotational kinetic energy of the diatomic molecule are diagonal in the SF total angular momentum coupled basis 
\begin{equation}\label{RotationalMatrix}
\langle  JMlj(NS)| B_e \hat{N}^2 |J'M'l'j'(N'S')\rangle
=B_e N(N+1) \delta_{JJ’}\delta_{MM’}\delta_{ll’} \delta_{j j’}\delta_{N N’} \delta_{S S’}
\end{equation} 
as are the matrix elements of the spin-rotation matrix interaction (since our angular basis functions are the eigenstates of $\hat{j}^2$ 
\begin{multline}\label{SpinRotationMatrix}
 \langle JMlj(NS)| \gamma_\text{sr} \hat{\mathbf{N}} \cdot \hat{\mathbf{S}} |J'M'l'j'(N'S')\rangle
=\delta_{JJ’}\delta_{MM’}\delta_{ll’} \delta_{j j’} \delta_{N N’} \delta_{S S’} \\ \times  \frac{1}{2}\gamma_\text{sr}  [j(j+1)-N(N+1)-S(S+1)]
\end{multline}

Finally, the matrix elements of the atom-molecule interaction potential are diagonal in $J$ as a consequence of rotational invariance of intermolecular interactions
\begin{multline}\label{InteractionPotentialMatrix}
\langle JMlj(NS) |V(R,\theta) |J'M'l'j'(N'S')\rangle
=\delta_{JJ’}\delta_{MM’} \delta_{S S’}(-1)^{J+S+2j’+\lambda}
[(2j’+1)(2j+1) \\ \times (2N+1)(2N’+1)(2l+1)(2l’+1)]^{1/2}
\sum_\lambda V_\lambda(R) 
\begin{Bmatrix}
N&j&S\\
j’& N’ &\lambda
\end{Bmatrix}
\begin{Bmatrix}
j&l&J\\
l’& j’ &\lambda
\end{Bmatrix}
\begin{pmatrix}
N&\lambda&N’\\
0& 0 &0
\end{pmatrix}
\begin{pmatrix}
l&\lambda&l’\\
0& 0 &0
\end{pmatrix},
\end{multline}
where $V_\lambda(R)$ are the radial expansion coefficients of the atom-molecule interaction PES in Legendre polynomials, $V(R,\theta)=\sum_\lambda V_\lambda(R)P_\lambda(\cos\theta)$. 

\subsection{Unphysical states in the SF representation and their elimination}



 Previous theoretical work has focused on the unphysical states arising in the BF formulation of the collision problem \cite{Tscherbul:10}. To gain additional insight into the origin of unphysical states, here we reformulate the problem in the SF total angular momentum representation.  As shown below, this reformulation makes explicit the symmetry of the asymptotic Hamiltonian related to to the conservation of the squared orbital angular momentum of the collision complex $\hat{L}^2$. This allows us to identify the basis states responsible for the formation of unphysical states and to suggest a general strategy for eliminating these states from quantum scattering calculations on molecular collisions in external fields. 
  
  Consider the matrix of the asymptotic Hamiltonian (\ref{ZeemanMatrix}) in the SF total angular momentum representation.   
  Restricting our attention for the moment to the ground rotational state ($N=0$, $j=1/2$), we choose a minimal  basis consisting of two total angular momentum states, $J=1/2$ and $J=3/2$. The corresponding values of $l$ range from $|j-J|$ to $|j+J|$. Using the shorthand notation $\ket{J l j}$  for the  basis states $\ket{Jlj(NS)}$ within the $N=0$ and $S=1/2$ subspace of interest, there are two  basis states, $\ket{\frac{1}{2}0 \frac{1}{2}}$ and $\ket{\frac{1}{2}1 \frac{1}{2}}$ for  $J=1/2$ and two basis states, $\ket{\frac{3}{2} 1\frac{1}{2}}$ and $\ket{\frac{3}{2} 2\frac{1}{2}}$ for $J=3/2$. The asymptotic Hamiltonian is represented in this basis by a $4\times 4$ matrix (with the basis functions arranged in the sequence    $\ket{\frac{1}{2}0 \frac{1}{2}}$, $\ket{\frac{1}{2}1 \frac{1}{2}}$, $\ket{\frac{3}{2} 1\frac{1}{2}}$, $\ket{\frac{3}{2} 2\frac{1}{2}}$)
 \begin{equation}\label{Hmol_N0}
     \hat{H}_{mol} = \begin{bmatrix}
     \mu_{0}B & 0 & 0 & 0 \\
     0 & -\frac{1}{3}\mu_{0}B & \frac{2\sqrt{2}}{3}\mu_{0}B & 0 \\
     0 & \frac{2\sqrt{2}}{3}\mu_{0}B & \frac{1}{3}\mu_{0}B & 0 \\
     0 & 0 & 0 & -\frac{1}{5}\mu_{0}B
     \end{bmatrix}
 \end{equation} 
 This matrix is diagonal in $l$ as a consequence of $\hat{L}^2$ commuting with $\hat{H}_\text{mol}$, and thus the matrix representation (\ref{Hmol_N0}) is notably simpler than that of $\hat{H}_\text{mol}$ in the BF total angular momentum basis \cite{Tscherbul:10} (where it is given by a full matrix). The block-diagonal structure of the matrix permits analytical diagonalization to yield the eigenvalues
 \begin{equation}\
     \lambda_{1} = \lambda_2 = \mu_0 B,\enspace \lambda_{3} = -\mu_0 B, \enspace \lambda_{4}= -\frac{1}{5}\mu_0 B
 \end{equation}
and the corresponding eigenvectors 
\begin{align}\label{eigenvec_N0}
    \ket{1} &= \ket{J=1/2,l=0,j=1/2} \notag \\
    \ket{2} &= \frac{1}{\sqrt{3}} \ket{J=1/2,l=1,j=1/2} + \sqrt{\frac{2}{3}} \ket{J=3/2,l=1,j=1/2} \notag \\
    \ket{3} &= -\sqrt{\frac{2}{3}} \ket{J=1/2,l=1,j=1/2} + \frac{1}{\sqrt{3}} \ket{J=3/2,l=1,j=1/2} \notag \\
    \ket{4} &= \ket{J=3/2,l=2,j=1/2}
\end{align}
 We observe that three of the eigenvalues correspond to the physical Zeeman states of a rotationless $^2\Sigma$ molecule in a magnetic field ($\pm \mu_0 B $), whereas the eigenvalue $\lambda_4$ is unphysical. By examining the $1\times 1$ subblock of the asymptotic Hamiltonian matrix in Eq.~(\ref{Hmol_N0}) that corresponds to  $l=2$, we notice that it  is spanned by a single basis state $\ket{\frac{3}{2} 2\frac{1}{2}}$ with $\lambda_4=\bra{\frac{3}{2} 2\frac{1}{2}} \hat{H}_\text{mol} \ket{\frac{3}{2} 2\frac{1}{2}}$. Significantly, this is not a complete basis for $l=2$: the state $\ket{\frac{5}{2} 2\frac{1}{2}}$ is missing because the SF basis set $\ket{Jlj(NS)}$  defined above only includes (by construction) the two lowest values of  $J=1/2$ and $3/2$.  This is reflected in the absence of the contribution from basis state $|\frac{5}{2}2\frac{1}{2}\rangle$ in the eigenvector  $|4\rangle$ [Eq.~(\ref{eigenvec_N0})].
It is now clear that the unphysical state $\ket{4}$ arises from the incompleteness of the $l=2$  basis set caused by the truncation procedure, which limits the range of available $J$ states to 1/2 and 3/2.

 The above arguments suggest that, in order to  eliminate unphysical states from the spectrum of the asymptotic Hamiltonian, we can simply remove the unphysical $1\times 1$ subblock corresponding to $l=2$ in Eq.~(\ref{Hmol_N0}) by {excluding} the state  $\ket{\frac{3}{2} 2\frac{1}{2}}$ from the basis set. Alternatively, we could {complete} the $l=2$ subblock by {including} the basis state $\ket{\frac{5}{2} 2\frac{1}{2}}$.  The former procedure simply eliminates the unphysical eigenvalue $\lambda_4$, whereas the latter results in the following augmented matrix of the asymptotic Hamiltonian   (with the basis functions arranged as $\ket{\frac{1}{2}0 \frac{1}{2}}$, $\ket{\frac{1}{2}1 \frac{1}{2}}$, $\ket{\frac{3}{2} 1\frac{1}{2}}$, $\ket{\frac{3}{2} 2\frac{1}{2}}$, $\ket{\frac{5}{2} 2\frac{1}{2}}$)
\begin{equation}
     \hat{H}_{mol} = \begin{bmatrix}
     \mu_{0}B & 0 & 0 & 0 & 0 \\
     0 & -\frac{1}{3}\mu_{0}B & \frac{2\sqrt{2}}{3}\mu_{0}B & 0 & 0 \\
     0 & \frac{2\sqrt{2}}{3}\mu_{0}B & \frac{1}{3}\mu_{0}B & 0 & 0 \\
     0 & 0 & 0 & -\frac{1}{5}\mu_{0}B & \frac{2\sqrt{6}}{5} \mu_{0}B \\
     0 & 0 & 0 & \frac{2\sqrt{6}}{5}\mu_{0}B & \frac{1}{5}\mu_{0}B \\
     \end{bmatrix}
 \end{equation} 
The  $2\times 2$  block corresponding to $l=2$ is now complete, and its eigenvalues  are
 \begin{equation}
     \lambda_{4} = \mu_0 B, \enspace \lambda_{5}= -\mu_0 B
 \end{equation}
 with the corresponding eigenvectors   
\begin{align}
    \ket{4} &= \sqrt{\frac{2}{5}}\ket{J=3/2,l=2,j=1/2} + \sqrt{\frac{3}{5}}\ket{J=5/2,l=2,j=1/2} \notag \\
    \ket{5} &= -\sqrt{\frac{3}{5}}\ket{J=3/2,l=2,j=1/2} + \sqrt{\frac{2}{5}}\ket{J=5/2,l=2,j=1/2}
\end{align}
We observe that {the unphysical states have disappeared as expected, since the matrix representation of the asymptotic Hamiltonian in the $l=2$ sector is now complete.}

\begin{figure}[t]
\begin{center}
\includegraphics[height=0.25\textheight,keepaspectratio]{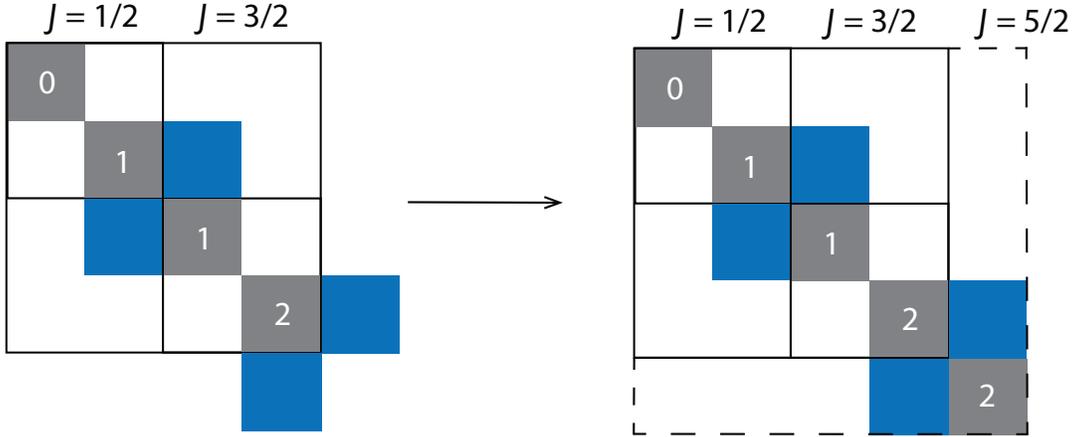}
\end{center}
\caption{Schematic illustration of the basis set completion procedure used to construct the augmented SF basis. (Left): Matrix representation of the asymptotic Hamiltonian in the SF total angular momentum basis truncated using the standard procedure. Only the $N=0$ ($j=1/2$) basis states are included, giving rise to two $l$ basis states for each $J$ with $l=J\pm 1/2$. The Zeeman Hamiltonian is diagonal in $l$ but off-diagonal in $J$ (see text). The values of $J$ corresponding to each basis function are indicated at the top of each block;  the diagonal blocks are labeled by the orbital angular momentum quantum number $l$. The diagonal (off-diagonal) magnetic field couplings are represented by grey (blue) squares. (Right): Matrix representation  of the asymptotic Hamiltonian in the augmented SF basis including the $l=2,J=5/2$ state required to obtain the  complete angular momentum  basis set  for $l=2$. No unphysical states occur in the augmented SF representation.  }
\label{fig:Extension}
\end{figure}

We now generalize the above discussion to the case of arbitrary SF basis sets $\ket{JMlj(NS)} $ with the goal of devising a general procedure for eliminating unphysical Zeeman states. 
As we have just seen, the origin of these states can be attributed to the incompleteness of the $\ket{JMlj(NS)} $ basis truncated at a finite $J$ that is below the maximum value allowed by the angular momentum addition rules. As shown in Fig.~\ref{fig:Extension},  conventional truncation of the basis set at a predetermined cutoff value of $J=J_\text{max}$ leaves out the important field-induced couplings between the states of  different $J$ but same $l$, thereby producing unphysical Zeeman states.
 Therefore, in order to eliminate the unphysical states, we need to remove the ``dangling'' field couplings.


Two general strategies for achieving this are immediately apparent.  First, as observed above, one can exclude the  $\ket{JMlj(NS)}$ basis states with the $l$ values, for which $j+l\ge J_\text{max}$. Test calculations show that the exclusion strategy does not produce meaningful results, which suggests that the excluded states play a non-negligible role in collision dynamics.
An alternative idea, illustrated in Fig.~\ref{fig:Extension},  is to {\it complete} the SF basis for each $l$ by {\it including} the full range of  $J$ values up to $J_\text{max}(j,l)=j+l$.  We will refer to such a basis as the {\it augmented SF basis}. We note that the {completion procedure} can quickly become computationally expensive because  $J_\text{max}(j,l)=j+l$ can become very large  for high $j$, thereby defeating the purpose of using the total angular momentum representation  (which saves computational time by including a {limited number} of $J$ states with $J<J_\text{max}$) \cite{Tscherbul:10}.  As a result, the completion procedure should be limited to low $N$ and $j$ values (i.e., those in the vicinity of the collision threshold) to prevent rapid growth of the basis set. 
  
  
  To implement the basis set completion procedure in practice, we proceed to define and truncate the augmented SF total angular momentum basis using the following algorithm, assuming that the values of molecular spin ($S=1/2$ for $^2\Sigma$ molecules) and the upper bounds  $N_\text{max}$ and  $J_\text{max}$ are supplied as input parameters
\begin{algorithmic}
\State $N_b \gets 0$
 \While{$N\leq N_\text{max}$} 
 
\If{$N\leq N_\text{corr}$} 
   \State $l_\text{max} \gets J_\text{max} + N + S - 1 $
 
  \While{$|N-S|\leq j\leq N+S$}   
  
    \While{$0\leq l\leq l_\text{max}$}  
    
    \While{$\max(|M|,|l-j|)\leq J \leq l + j$}
    
    \State Add state  $\ket{Jlj(NS)}$ to the basis
   \State $N_b+1  \gets N_b  $
   
       \EndWhile
   
    \EndWhile
 
 \EndWhile
 
\EndIf 
   \EndWhile
\end{algorithmic}

The outer loop runs over the rotational quantum numbers $N$ and the first if statement ensures that the augmented SF basis is constructed only for  a subset of $N$ values of interest ($N\le N_\text{corr}$,  where $N_\text{corr}$ is the number of channels, for which unphysical states are to be eliminated, e.g., near-threshold or open channels).  In the next step, we  define the maximum orbital angular momentum quantum number $\l_\text{max}$ such that it does not exceed the maximum possible value allowed by angular momentum addition rules ($J_\text{max}+N+S$). Finally, we construct a loop over the total angular momentum quantum number $J$  in such a way as to ensure basis set completeness for each $l$ as described above.

\color{black}

\section{Results and discussion}

In the preceding section, we have described a basis set completion procedure to construct the augmented SF total angular momentum representation, which is free from  unphysical states. Here, we illustrate the procedure by calculating the spectrum of the asymptotic Hamiltonian of the CaH molecule in a magnetic field. We also test the accuracy of  the augmented SF basis by carrying out quantum scattering calculations on  cold He~+~CaH collisions in a magnetic field.

\subsection{Unphysical state-free representation: Basis set completion procedure and augmented SF total angular momentum representation}

\begin{figure}[t]
\begin{center}
\includegraphics[height=0.35\textheight,keepaspectratio]{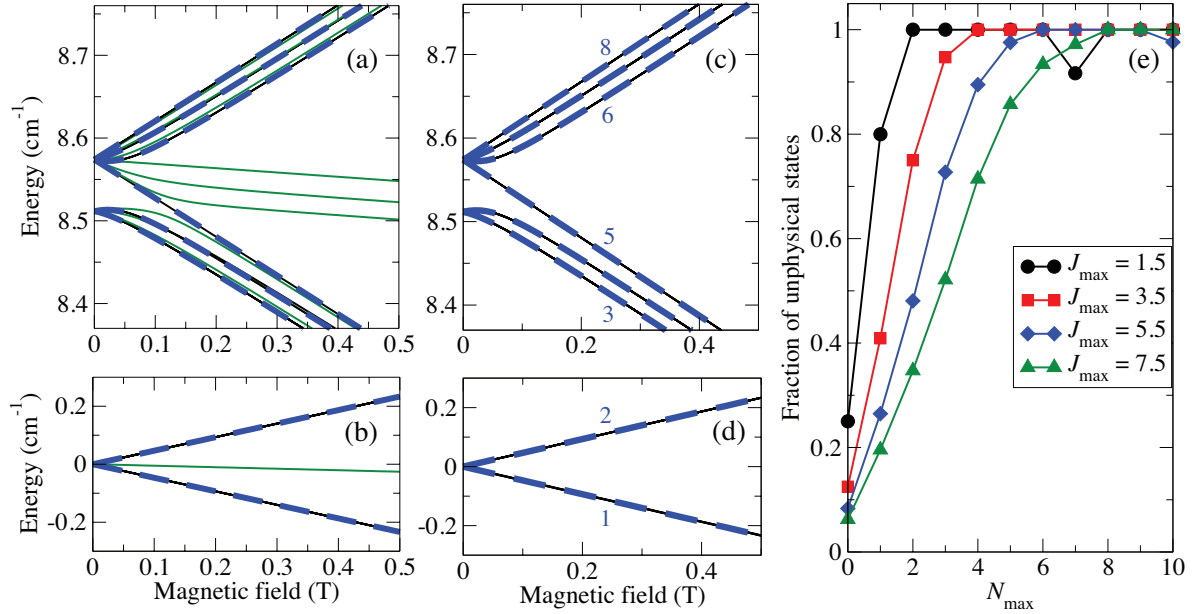}
\end{center}
\caption{Eigenvalues of the asymptotic Hamiltonian for CaH$(^2\Sigma^+)$ in the $N=1$ (a) and $N=0$ (b) rotational states plotted as a function of external magnetic field. The SF basis used to compute the eigenvalues includes all $J$ states  with $J\le J_\text{max}=7/2$. The physical Zeeman levels of CaH$(^2\Sigma^+)$  are shown by dashed lines; unphysical Zeeman levels are shown by green lines. (c), (d) Same as panels (a), (b) but computed using the augmented SF basis  (see text). Note the absence of unphysical states in the latter basis.}
\label{fig:levels}
\end{figure}

Figures~\ref{fig:levels}(a)-(d) show the eigenspectrum of the asymptotic Hamiltonian as a function of external magnetic field calculated for a prototypical $^2\Sigma$ molecule CaH($^{2}\Sigma^+$) using the standard SF basis set $\ket{JMlj(NS)}$ truncated at $J_\text{max}=7/2$. While there is only a single unphysical Zeeman state in the ground $N=0$ manifold, there are 9 such states in the $N=1 $ manifold (shown in Fig.~\ref{fig:levels}(a) by the green lines).  The fraction of unphysical states $n_u/(n_p+n_u)$, where $n_u$($n_p$) is the number of unphysical (physical) Zeeman levels, grows rapidly with $N$ as shown in Fig.~\ref{fig:levels}(e). Even though the growth is less steep for higher $J_\text{max}$,  we observe that most of the high-$N$ eigenstates of $\hat{H}_\text{mol}$ are unphysical for  $J_\text{max}\le 7.5$.


The physical Zeeman states in the strong field limit are well approximated by the fully uncoupled basis states $\ket{NM_NM_S}$ (see Sec.~II above), where $N$ is the rotational quantum number, and $M_N$ and $M_S$ are the projections of $\mathbf{N}$ and $\mathbf{S}$  onto the magnetic field axis. 

We note that the Zeeman sublevels $\ket{5}$ and $\ket{8}$ in the $N=1$ manifold are special in that they correspond to the fully spin-stretched states $\ket{1-1-\frac{1}{2}}$ and  $\ket{11\frac{1}{2}}$ that are are not affected by the spin-rotation interaction.
As such,  $M_S$ and $M_N$ remain rigorously good quantum numbers for these states   {\it at all magnetic fields}.  In contrast, the other $N=1$ Zeeman states are coupled by the spin-rotation interaction, resulting in substantial mixing between their $M_S$ and $M_N$ components.
For example,  $\ket{6}=0.94\ket{1-1\frac{1}{2}} + 0.33\ket{10-\frac{1}{2}}$ at $B=0.1$~T.
This mixing is large in CaH due to its large spin-rotation constant, and will generally be much weaker in heavier molecules  with smaller spin-rotation constants such as YbF \cite{Tscherbul:07} at the same magnetic field, resulting in purer $\ket{NM_NM_S}$ states.


Returning to the discussion of unphysical states,  we observe from  Fig.~\ref{fig:levels}(a), that some unphysical states can lie very close in energy to the true physical  states (such as state $\ket{8}$).  As stated in the Introduction, this could lead to difficulties in calculating state-to-state scattering cross sections and near-threshold bound states of atom-molecule complexes in external fields,  motivating the search for an unphysical state-free representation of the asymptotic Hamiltonian.  

In Figs.~\ref{fig:levels}(c) and (d) we plot the spectrum of the asymptotic Hamiltonian calculated using the augmented SF total angular momentum basis for $N=0$ and 1 constructed as described in Sec. IIB.  We observe that the spectrum is completely free of unphysical states, as expected for a complete angular momentum basis set (see the previous section). This shows that the augmented SF basis provides an appealing, unphysical states-free representation of the asymptotic Hamiltonian.






\subsection{State-to-state He-CaH scattering in a magnetic field}

In this section, we will validate the augmented SF total angular momentum basis proposed in Sec.~IIB. 
To this end, we compare state-to-state cross sections for  $^{40}$He~+~$^{40}$CaH collisions  calculated using the new basis  with benchmark results obtained using the fully uncoupled SF representation \cite{Volpi:02,Krems:04}. 
{To obtain converged solutions of the time-independent  Schr\"odinger equation (\ref{eq:Heff}) we propagate the CC equations on the radial grid from $R=2\,a_0$ to $R=60 a_0$ with the grid step of $0.04\, a_0$, which produces the integral cross sections converged to $<$5\%. The total angular momentum basis sets included all states with $N\le 5$ and $J\le 6.5$. We use a highly accurate {\it ab initio} He-CaH PES developed in  Ref.~\cite{Groenenboom:03} and employed in a number of low-temperature scattering calculations \cite{Krems:03,Balakrishnan:03, Tscherbul:06,Tscherbul:06b,Abrahamsson:07,Tscherbul:10}. The PES is expanded in Legendre polynomials as $V(R,\theta)=\sum_{\lambda=0}^{12}V_\lambda(R)P_\lambda(\cos\theta)$ and the radial coefficients $V_\lambda(R)$ in Eq.~(\ref{InteractionPotentialMatrix}) are evaluated on a 40-point Gauss-Legendre quadrature. The convergence parameters used in the benchmark calculations are the same as (or more extensive than) those used in our previous work \cite{Tscherbul:10}.}

\begin{figure}[t]
\begin{center}
\includegraphics[height=0.4\textheight,keepaspectratio]{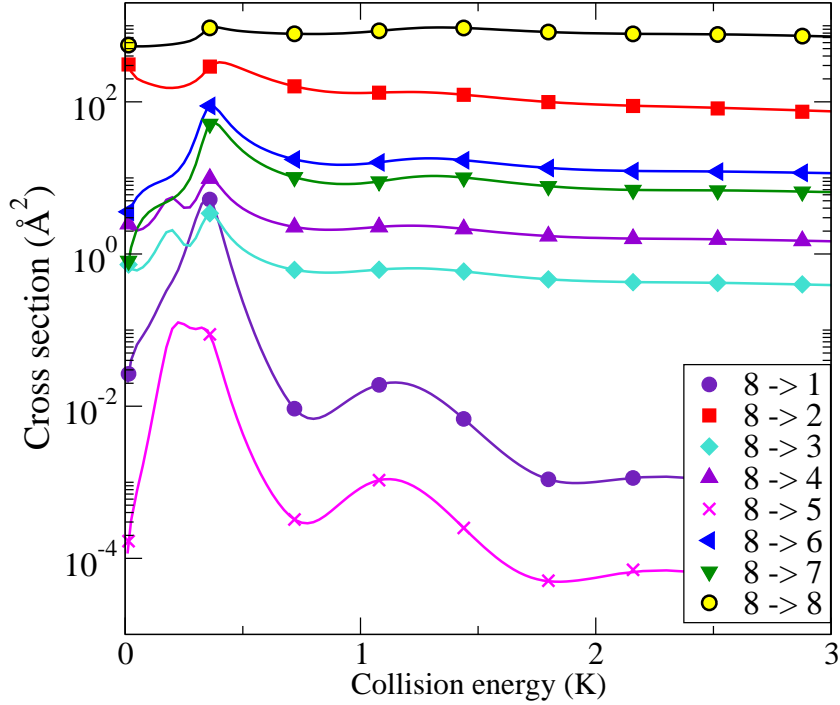}
\end{center}
\caption{Collision energy dependence of state-to-state integral cross sections for cold $^4$He~+~$^{40}$CaH collisions in a magnetic field of 0.1~T  with $^{40}$CaH molecules initially in the fully spin-stretched rotationally excited state $\ket{8}=  \ket{11\frac{1}{2}}$. Symbols -- present calculations using the augmented SF total angular momentum basis with $J_\text{max}=6.5$.  Full lines -- benchmark calculations based on the uncoupled space-fixed representation \cite{Krems:04,Tscherbul:06}. See Figs.~2(c)-(d) for the Zeeman diagram of CaH showing the final states  $\ket{1}$ - $\ket{8}$.} 
\label{fig:xs}
\end{figure}

Figure~\ref{fig:xs} shows the collision energy dependence of state-to-state cross sections for  $^{40}$He~+~$^{40}$CaH collisions  calculated using the augmented SF total angular momentum basis vs. the benchmark values  \cite{Volpi:02,Krems:04}.
The cross sections calculated using two completely unrelated basis sets are in excellent agreement with each other for all final states ($\ket{1}$ - $\ket{8}$) and collision energies. Significantly, the agreement does not deteriorate near scattering resonances near 0.5~K, which are very sensitive to fine details of collision dynamics.  These results strongly suggest that the augmented SF total angular momentum representation provides an accurate description of quantum state-to-state collision dynamics.


 As shown  previously  \cite{Tscherbul:06} collisions with cold He atoms result in rapid rotational relaxation of CaH molecules, which transfers molecular population from the rotationally excited initial state  $\ket{8}$  to lower-lying rotational and Zeeman levels (we assume that the collision energy is small enough so that rotational excitation of CaH into higher rotational manifolds ($N\ge 2$) is energetically forbidden).
 
The state-to-state cross sections  for rotational-Zeeman relaxation of $^{40}$CaH molecules in a cold $^{4}$He buffer gas follow a clear trend.
The largest cross sections are those for spin-conserving rotational relaxation $\ket{8}\to \ket{2}$ or $\ket{11\frac{1}{2}} \to \ket{00\frac{1}{2}}$. The second most efficient are the processes $\ket{8} \to \ket{6}$ ($\ket{11\frac{1}{2}} \to \ket{1-1\frac{1}{2}}$) and $\ket{8}\to \ket{7}$ ($\ket{11\frac{1}{2}} \to \ket{10\frac{1}{2}}$) that conserve $N$ and $M_S$ but change $M_N$.
The slowest transitions are those which conserve $N$ but change $M_S$: $\ket{8}\to \ket{5}$ ($\ket{11\frac{1}{2}} \to \ket{1-1-\frac{1}{2}}$), $\ket{8}\to \ket{4}$ ($\ket{11\frac{1}{2}} \to \ket{10-\frac{1}{2}}$), and, to a less extent, $\ket{8}\to \ket{3}$ ($\ket{11\frac{1}{2}} \to \ket{11-\frac{1}{2}}$). 

Remarkably, as shown in Fig.~\ref{fig:xs}, the slowest relaxation transition $\ket{8}\to \ket{5}$  occurs between  the pure spin-stretched states $\ket{11\frac{1}{2}} $ and $\ket{1-1-\frac{1}{2}}$. 
Note that all other $N=1$ states ($\ket{3}$, $\ket{4}$, $\ket{6}$, and $\ket{7}$) are not pure  $\ket{NM_NM_S}$ states, i.e., they have a non-negligible admixture of the $\ket{N'M_N'M_S'}$ components with $M_N'\ne M_N$ and $M_S'\ne M_S$ due to the spin-rotation interaction (see above). As a result, transitions originating from state $\ket{8}$ to these ``unpure'' final states   are not suppressed compared to the $\ket{8}\to \ket{5}$ transition.

 \color{black}

 
 \section{Summary and conclusions}
 
 We have  demonstrated the possibility of computationally efficient quantum dynamical calculations of state-to-state scattering cross sections for cold molecular collisions in an external magnetic field. Prior to this work, such calculations have been hampered by the presence of unphysical states in the eigenspectrum of the asymptotic Hamiltonian in the total angular momentum representation \cite{Tscherbul:10,Tscherbul:11,Suleimanov:12,Morita:18}. We have developed a general procedure for eliminating the unphysical states 
 based on the observation that the asymptotic Hamiltonian  commutes with the  orbital angular momentum squared ($\hat{{L}}^2$) of the collision complex. As a result, the asymptotic Hamiltonian is diagonal in the basis of simultaneous eigenstates of $\hat{L}^2$ and $\hat{J}^2$, i.e., the SF total angular momentum eigenstates.  Using this property, we determined that the origin of unphysical states in molecular Zeeman spectra  can be traced back to the incompleteness of fixed-$l$ angular momentum basis sets, which do not  have a sufficient number of $J$-states ranging from $|j-l|$ to $j+l$, leading in turn to an incomplete representation of the asymptotic Hamiltonian. 
 
 Armed with this insight, we developed a general procedure for removing the unphysical states from the spectrum of the asymptotic Hamiltonian. The procedure extends the number of $J$ states for each $l$ to include the complete set of quantum numbers ranging from $J=|j-l|$ to $J=J+l$ as required by the angular momentum addition rules. This procedure can be implemented  for each individual rotational state in the basis set, and ensures  the completeness of the SF total angular momentum basis for each $l$, thereby keeping only the physical eigenstates in the spectrum. We have verified, through rigorous quantum scattering calculations of cold $^4$He~+~$^{40}$CaH collisions, that the augmented SF total angular momentum basis provides accurate state-to-state scattering cross sections for all final states of $^{40}$CaH over the range of collision energies  0.01 -- 3~K. 

 The augmented SF total angular momentum basis proposed in this work  removes the congestion caused by unphysical Zeeman states  in  the spectrum of the asymptotic Hamiltonian, thereby facilitating quantum scattering calculations of state-to-state cross sections and possibly near-threshold energy levels of strongly anisotropic  atom-molecule collision complexes in the presence of an external  magnetic field. Rigorous calculations of such spectra  play a central role in the identification and assignment of magnetic Feshbach resonances in ultracold atom-molecule  \cite{Tscherbul:07,Hummon:11,Wang:21} and molecule-molecule \cite{Tscherbul:09c,Suleimanov:11} collisions.
  
 Because our  procedure relies on the conservation of the orbital angular momentum of the collision complex in the limit $R\to \infty$,  it is not directly applicable to the BF total angular momentum representation, where the operator $\hat{L}^2$ is not diagonal.     In future work, it would be desirable to extend our procedure to eliminate the unphysical states directly in the BF representation. It would also be interesting to apply the procedure  to more complex molecular states of, e.g., $^3\Sigma$  and $^2\Pi$ symmetries and to molecular collisions in the presence of an external electric field \cite{Tscherbul:12}.
 
 \section*{Acknowledgements}
 This work was supported by the NSF (Grant No. PHY-1912668).
 
 \section*{Data availability statement}
The data that support the findings of this study are available from the corresponding author upon reasonable request.

\bibliography{cold_mol}

\begin{thebibliography}{48}%
\makeatletter
\providecommand \@ifxundefined [1]{%
 \@ifx{#1\undefined}
}%
\providecommand \@ifnum [1]{%
 \ifnum #1\expandafter \@firstoftwo
 \else \expandafter \@secondoftwo
 \fi
}%
\providecommand \@ifx [1]{%
 \ifx #1\expandafter \@firstoftwo
 \else \expandafter \@secondoftwo
 \fi
}%
\providecommand \natexlab [1]{#1}%
\providecommand \enquote  [1]{``#1''}%
\providecommand \bibnamefont  [1]{#1}%
\providecommand \bibfnamefont [1]{#1}%
\providecommand \citenamefont [1]{#1}%
\providecommand \href@noop [0]{\@secondoftwo}%
\providecommand \href [0]{\begingroup \@sanitize@url \@href}%
\providecommand \@href[1]{\@@startlink{#1}\@@href}%
\providecommand \@@href[1]{\endgroup#1\@@endlink}%
\providecommand \@sanitize@url [0]{\catcode `\\12\catcode `\$12\catcode
  `\&12\catcode `\#12\catcode `\^12\catcode `\_12\catcode `\%12\relax}%
\providecommand \@@startlink[1]{}%
\providecommand \@@endlink[0]{}%
\providecommand \url  [0]{\begingroup\@sanitize@url \@url }%
\providecommand \@url [1]{\endgroup\@href {#1}{\urlprefix }}%
\providecommand \urlprefix  [0]{URL }%
\providecommand \Eprint [0]{\href }%
\providecommand \doibase [0]{https://doi.org/}%
\providecommand \selectlanguage [0]{\@gobble}%
\providecommand \bibinfo  [0]{\@secondoftwo}%
\providecommand \bibfield  [0]{\@secondoftwo}%
\providecommand \translation [1]{[#1]}%
\providecommand \BibitemOpen [0]{}%
\providecommand \bibitemStop [0]{}%
\providecommand \bibitemNoStop [0]{.\EOS\space}%
\providecommand \EOS [0]{\spacefactor3000\relax}%
\providecommand \BibitemShut  [1]{\csname bibitem#1\endcsname}%
\let\auto@bib@innerbib\@empty
\bibitem [{\citenamefont {Bohn}\ \emph {et~al.}(2017)\citenamefont {Bohn},
  \citenamefont {Rey},\ and\ \citenamefont {Ye}}]{Bohn:17}%
  \BibitemOpen
  \bibfield  {author} {\bibinfo {author} {\bibfnamefont {J.~L.}\ \bibnamefont
  {Bohn}}, \bibinfo {author} {\bibfnamefont {A.~M.}\ \bibnamefont {Rey}},\ and\
  \bibinfo {author} {\bibfnamefont {J.}~\bibnamefont {Ye}},\ }\bibfield
  {title} {\bibinfo {title} {Cold molecules: Progress in quantum engineering of
  chemistry and quantum matter},\ }\href@noop {} {\bibfield  {journal}
  {\bibinfo  {journal} {Science}\ }\textbf {\bibinfo {volume} {357}},\ \bibinfo
  {pages} {1002} (\bibinfo {year} {2017})}\BibitemShut {NoStop}%
\bibitem [{\citenamefont {Balakrishnan}\ and\ \citenamefont
  {Dalgarno}(2001)}]{Balakrishnan:01}%
  \BibitemOpen
  \bibfield  {author} {\bibinfo {author} {\bibfnamefont {N.}~\bibnamefont
  {Balakrishnan}}\ and\ \bibinfo {author} {\bibfnamefont {A.}~\bibnamefont
  {Dalgarno}},\ }\bibfield  {title} {\bibinfo {title} {Chemistry at ultracold
  temperatures},\ }\href@noop {} {\bibfield  {journal} {\bibinfo  {journal}
  {Chem. Phys. Lett.}\ }\textbf {\bibinfo {volume} {341}},\ \bibinfo {pages}
  {652} (\bibinfo {year} {2001})}\BibitemShut {NoStop}%
\bibitem [{\citenamefont {Tscherbul}\ and\ \citenamefont
  {Krems}(2015)}]{Tscherbul:15}%
  \BibitemOpen
  \bibfield  {author} {\bibinfo {author} {\bibfnamefont {T.~V.}\ \bibnamefont
  {Tscherbul}}\ and\ \bibinfo {author} {\bibfnamefont {R.~V.}\ \bibnamefont
  {Krems}},\ }\bibfield  {title} {\bibinfo {title} {Tuning bimolecular chemical
  reactions by electric fields},\ }\href
  {https://doi.org/10.1103/PhysRevLett.115.023201} {\bibfield  {journal}
  {\bibinfo  {journal} {Phys. Rev. Lett.}\ }\textbf {\bibinfo {volume} {115}},\
  \bibinfo {pages} {023201} (\bibinfo {year} {2015})}\BibitemShut {NoStop}%
\bibitem [{\citenamefont {Hermsmeier}\ \emph {et~al.}(2021)\citenamefont
  {Hermsmeier}, \citenamefont {K\l{}os}, \citenamefont {Kotochigova},\ and\
  \citenamefont {Tscherbul}}]{Hermsmeier:21}%
  \BibitemOpen
  \bibfield  {author} {\bibinfo {author} {\bibfnamefont {R.}~\bibnamefont
  {Hermsmeier}}, \bibinfo {author} {\bibfnamefont {J.}~\bibnamefont {K\l{}os}},
  \bibinfo {author} {\bibfnamefont {S.}~\bibnamefont {Kotochigova}},\ and\
  \bibinfo {author} {\bibfnamefont {T.~V.}\ \bibnamefont {Tscherbul}},\
  }\bibfield  {title} {\bibinfo {title} {Quantum spin state selectivity and
  magnetic tuning of ultracold chemical reactions of triplet alkali-metal
  dimers with alkali-metal atoms},\ }\href
  {https://doi.org/10.1103/PhysRevLett.127.103402} {\bibfield  {journal}
  {\bibinfo  {journal} {Phys. Rev. Lett.}\ }\textbf {\bibinfo {volume} {127}},\
  \bibinfo {pages} {103402} (\bibinfo {year} {2021})}\BibitemShut {NoStop}%
\bibitem [{\citenamefont {Croft}\ and\ \citenamefont {Bohn}(2014)}]{Croft:14}%
  \BibitemOpen
  \bibfield  {author} {\bibinfo {author} {\bibfnamefont {J.~F.~E.}\
  \bibnamefont {Croft}}\ and\ \bibinfo {author} {\bibfnamefont {J.~L.}\
  \bibnamefont {Bohn}},\ }\bibfield  {title} {\bibinfo {title} {Long-lived
  complexes and chaos in ultracold molecular collisions},\ }\href
  {https://doi.org/10.1103/PhysRevA.89.012714} {\bibfield  {journal} {\bibinfo
  {journal} {Phys. Rev. A}\ }\textbf {\bibinfo {volume} {89}},\ \bibinfo
  {pages} {012714} (\bibinfo {year} {2014})}\BibitemShut {NoStop}%
\bibitem [{\citenamefont {Liu}\ \emph {et~al.}(2020)\citenamefont {Liu},
  \citenamefont {Hu}, \citenamefont {Nichols}, \citenamefont {Grimes},
  \citenamefont {Karman}, \citenamefont {Guo},\ and\ \citenamefont
  {Ni}}]{Liu:20}%
  \BibitemOpen
  \bibfield  {author} {\bibinfo {author} {\bibfnamefont {Y.}~\bibnamefont
  {Liu}}, \bibinfo {author} {\bibfnamefont {M.-G.}\ \bibnamefont {Hu}},
  \bibinfo {author} {\bibfnamefont {M.~A.}\ \bibnamefont {Nichols}}, \bibinfo
  {author} {\bibfnamefont {D.~D.}\ \bibnamefont {Grimes}}, \bibinfo {author}
  {\bibfnamefont {T.}~\bibnamefont {Karman}}, \bibinfo {author} {\bibfnamefont
  {H.}~\bibnamefont {Guo}},\ and\ \bibinfo {author} {\bibfnamefont {K.-K.}\
  \bibnamefont {Ni}},\ }\bibfield  {title} {\bibinfo {title} {Photo-excitation
  of long-lived transient intermediates in ultracold reactions},\ }\href
  {https://doi.org/10.1038/s41567-020-0968-8} {\bibfield  {journal} {\bibinfo
  {journal} {Nat. Phys.}\ }\textbf {\bibinfo {volume} {16}},\ \bibinfo {pages}
  {1132} (\bibinfo {year} {2020})}\BibitemShut {NoStop}%
\bibitem [{\citenamefont {Devolder}\ \emph {et~al.}(2021)\citenamefont
  {Devolder}, \citenamefont {Brumer},\ and\ \citenamefont
  {Tscherbul}}]{Devolder:21}%
  \BibitemOpen
  \bibfield  {author} {\bibinfo {author} {\bibfnamefont {A.}~\bibnamefont
  {Devolder}}, \bibinfo {author} {\bibfnamefont {P.}~\bibnamefont {Brumer}},\
  and\ \bibinfo {author} {\bibfnamefont {T.~V.}\ \bibnamefont {Tscherbul}},\
  }\bibfield  {title} {\bibinfo {title} {Complete quantum coherent control of
  ultracold molecular collisions},\ }\href
  {https://doi.org/10.1103/PhysRevLett.126.153403} {\bibfield  {journal}
  {\bibinfo  {journal} {Phys. Rev. Lett.}\ }\textbf {\bibinfo {volume} {126}},\
  \bibinfo {pages} {153403} (\bibinfo {year} {2021})}\BibitemShut {NoStop}%
\bibitem [{\citenamefont {Bilitewski}\ \emph {et~al.}(2021)\citenamefont
  {Bilitewski}, \citenamefont {De~Marco}, \citenamefont {Li}, \citenamefont
  {Matsuda}, \citenamefont {Tobias}, \citenamefont {Valtolina}, \citenamefont
  {Ye},\ and\ \citenamefont {Rey}}]{Bilitewski:21}%
  \BibitemOpen
  \bibfield  {author} {\bibinfo {author} {\bibfnamefont {T.}~\bibnamefont
  {Bilitewski}}, \bibinfo {author} {\bibfnamefont {L.}~\bibnamefont
  {De~Marco}}, \bibinfo {author} {\bibfnamefont {J.-R.}\ \bibnamefont {Li}},
  \bibinfo {author} {\bibfnamefont {K.}~\bibnamefont {Matsuda}}, \bibinfo
  {author} {\bibfnamefont {W.~G.}\ \bibnamefont {Tobias}}, \bibinfo {author}
  {\bibfnamefont {G.}~\bibnamefont {Valtolina}}, \bibinfo {author}
  {\bibfnamefont {J.}~\bibnamefont {Ye}},\ and\ \bibinfo {author}
  {\bibfnamefont {A.~M.}\ \bibnamefont {Rey}},\ }\bibfield  {title} {\bibinfo
  {title} {Dynamical generation of spin squeezing in ultracold dipolar
  molecules},\ }\href {https://doi.org/10.1103/PhysRevLett.126.113401}
  {\bibfield  {journal} {\bibinfo  {journal} {Phys. Rev. Lett.}\ }\textbf
  {\bibinfo {volume} {126}},\ \bibinfo {pages} {113401} (\bibinfo {year}
  {2021})}\BibitemShut {NoStop}%
\bibitem [{\citenamefont {Croft}\ \emph {et~al.}(2017)\citenamefont {Croft},
  \citenamefont {Makrides}, \citenamefont {Li}, \citenamefont {Petrov},
  \citenamefont {Kendrick}, \citenamefont {Balakrishnan},\ and\ \citenamefont
  {Kotochigova}}]{Croft:17}%
  \BibitemOpen
  \bibfield  {author} {\bibinfo {author} {\bibfnamefont {J.~F.~E.}\
  \bibnamefont {Croft}}, \bibinfo {author} {\bibfnamefont {C.}~\bibnamefont
  {Makrides}}, \bibinfo {author} {\bibfnamefont {M.}~\bibnamefont {Li}},
  \bibinfo {author} {\bibfnamefont {A.}~\bibnamefont {Petrov}}, \bibinfo
  {author} {\bibfnamefont {B.~K.}\ \bibnamefont {Kendrick}}, \bibinfo {author}
  {\bibfnamefont {N.}~\bibnamefont {Balakrishnan}},\ and\ \bibinfo {author}
  {\bibfnamefont {S.}~\bibnamefont {Kotochigova}},\ }\bibfield  {title}
  {\bibinfo {title} {Universality and chaoticity in ultracold {K~+~KRb}
  chemical reactions},\ }\href@noop {} {\bibfield  {journal} {\bibinfo
  {journal} {Nat. Commun.}\ }\textbf {\bibinfo {volume} {8}},\ \bibinfo {pages}
  {15897} (\bibinfo {year} {2017})}\BibitemShut {NoStop}%
\bibitem [{\citenamefont {DeMille}(2002)}]{DeMille:02}%
  \BibitemOpen
  \bibfield  {author} {\bibinfo {author} {\bibfnamefont {D.}~\bibnamefont
  {DeMille}},\ }\bibfield  {title} {\bibinfo {title} {Quantum computation with
  trapped polar molecules},\ }\href@noop {} {\bibfield  {journal} {\bibinfo
  {journal} {Phys. Lev. Lett.}\ }\textbf {\bibinfo {volume} {88}} (\bibinfo
  {year} {2002})}\BibitemShut {NoStop}%
\bibitem [{\citenamefont {Andr{\'e}}\ \emph {et~al.}(2006)\citenamefont
  {Andr{\'e}}, \citenamefont {DeMille}, \citenamefont {Doyle}, \citenamefont
  {Lukin}, \citenamefont {Maxwell}, \citenamefont {Rabl}, \citenamefont
  {Schoelkopf},\ and\ \citenamefont {Zoller}}]{Andre:06}%
  \BibitemOpen
  \bibfield  {author} {\bibinfo {author} {\bibfnamefont {A.}~\bibnamefont
  {Andr{\'e}}}, \bibinfo {author} {\bibfnamefont {D.}~\bibnamefont {DeMille}},
  \bibinfo {author} {\bibfnamefont {J.~M.}\ \bibnamefont {Doyle}}, \bibinfo
  {author} {\bibfnamefont {M.~D.}\ \bibnamefont {Lukin}}, \bibinfo {author}
  {\bibfnamefont {S.~E.}\ \bibnamefont {Maxwell}}, \bibinfo {author}
  {\bibfnamefont {P.}~\bibnamefont {Rabl}}, \bibinfo {author} {\bibfnamefont
  {R.~J.}\ \bibnamefont {Schoelkopf}},\ and\ \bibinfo {author} {\bibfnamefont
  {P.}~\bibnamefont {Zoller}},\ }\bibfield  {title} {\bibinfo {title} {A
  coherent all-electrical interface between polar molecules and mesoscopic
  superconducting resonators},\ }\href {https://doi.org/10.1038/nphys386}
  {\bibfield  {journal} {\bibinfo  {journal} {Nat. Phys.}\ }\textbf {\bibinfo
  {volume} {2}},\ \bibinfo {pages} {636} (\bibinfo {year} {2006})}\BibitemShut
  {NoStop}%
\bibitem [{\citenamefont {Yelin}\ \emph {et~al.}(2006)\citenamefont {Yelin},
  \citenamefont {Kirby},\ and\ \citenamefont {C\^ot\'e}}]{Yelin:06}%
  \BibitemOpen
  \bibfield  {author} {\bibinfo {author} {\bibfnamefont {S.~F.}\ \bibnamefont
  {Yelin}}, \bibinfo {author} {\bibfnamefont {K.}~\bibnamefont {Kirby}},\ and\
  \bibinfo {author} {\bibfnamefont {R.}~\bibnamefont {C\^ot\'e}},\ }\bibfield
  {title} {\bibinfo {title} {Schemes for robust quantum computation with polar
  molecules},\ }\href {https://doi.org/10.1103/PhysRevA.74.050301} {\bibfield
  {journal} {\bibinfo  {journal} {Phys. Rev. A}\ }\textbf {\bibinfo {volume}
  {74}},\ \bibinfo {pages} {050301} (\bibinfo {year} {2006})}\BibitemShut
  {NoStop}%
\bibitem [{\citenamefont {Gorshkov}\ \emph {et~al.}(2011)\citenamefont
  {Gorshkov}, \citenamefont {Manmana}, \citenamefont {Chen}, \citenamefont
  {Demler}, \citenamefont {Lukin},\ and\ \citenamefont {Rey}}]{Gorshkov:11}%
  \BibitemOpen
  \bibfield  {author} {\bibinfo {author} {\bibfnamefont {A.~V.}\ \bibnamefont
  {Gorshkov}}, \bibinfo {author} {\bibfnamefont {S.~R.}\ \bibnamefont
  {Manmana}}, \bibinfo {author} {\bibfnamefont {G.}~\bibnamefont {Chen}},
  \bibinfo {author} {\bibfnamefont {E.}~\bibnamefont {Demler}}, \bibinfo
  {author} {\bibfnamefont {M.~D.}\ \bibnamefont {Lukin}},\ and\ \bibinfo
  {author} {\bibfnamefont {A.~M.}\ \bibnamefont {Rey}},\ }\bibfield  {title}
  {\bibinfo {title} {Quantum magnetism with polar alkali-metal dimers},\ }\href
  {https://doi.org/10.1103/PhysRevA.84.033619} {\bibfield  {journal} {\bibinfo
  {journal} {Phys. Rev. A}\ }\textbf {\bibinfo {volume} {84}},\ \bibinfo
  {pages} {033619} (\bibinfo {year} {2011})}\BibitemShut {NoStop}%
\bibitem [{\citenamefont {Hazzard}\ \emph {et~al.}(2013)\citenamefont
  {Hazzard}, \citenamefont {Manmana}, \citenamefont {Foss-Feig},\ and\
  \citenamefont {Rey}}]{Hazzard:13}%
  \BibitemOpen
  \bibfield  {author} {\bibinfo {author} {\bibfnamefont {K.~R.~A.}\
  \bibnamefont {Hazzard}}, \bibinfo {author} {\bibfnamefont {S.~R.}\
  \bibnamefont {Manmana}}, \bibinfo {author} {\bibfnamefont {M.}~\bibnamefont
  {Foss-Feig}},\ and\ \bibinfo {author} {\bibfnamefont {A.~M.}\ \bibnamefont
  {Rey}},\ }\bibfield  {title} {\bibinfo {title} {Far-from-equilibrium quantum
  magnetism with ultracold polar molecules},\ }\href
  {https://doi.org/10.1103/PhysRevLett.110.075301} {\bibfield  {journal}
  {\bibinfo  {journal} {Phys. Rev. Lett.}\ }\textbf {\bibinfo {volume} {110}},\
  \bibinfo {pages} {075301} (\bibinfo {year} {2013})}\BibitemShut {NoStop}%
\bibitem [{\citenamefont {Volpi}\ and\ \citenamefont {Bohn}(2002)}]{Volpi:02}%
  \BibitemOpen
  \bibfield  {author} {\bibinfo {author} {\bibfnamefont {A.}~\bibnamefont
  {Volpi}}\ and\ \bibinfo {author} {\bibfnamefont {J.~L.}\ \bibnamefont
  {Bohn}},\ }\bibfield  {title} {\bibinfo {title} {Magnetic-field effects in
  ultracold molecular collisions},\ }\href@noop {} {\bibfield  {journal}
  {\bibinfo  {journal} {Phys. Rev. A}\ }\textbf {\bibinfo {volume} {65}},\
  \bibinfo {pages} {052712} (\bibinfo {year} {2002})}\BibitemShut {NoStop}%
\bibitem [{\citenamefont {Krems}\ and\ \citenamefont
  {Dalgarno}(2004)}]{Krems:04}%
  \BibitemOpen
  \bibfield  {author} {\bibinfo {author} {\bibfnamefont {R.~V.}\ \bibnamefont
  {Krems}}\ and\ \bibinfo {author} {\bibfnamefont {A.}~\bibnamefont
  {Dalgarno}},\ }\bibfield  {title} {\bibinfo {title} {Quantum-mechanical
  theory of atom-molecule and molecular collisions in a magnetic field: {Spin}
  depolarization},\ }\href {https://doi.org/10.1063/1.1636691} {\bibfield
  {journal} {\bibinfo  {journal} {J. Chem. Phys.}\ }\textbf {\bibinfo {volume}
  {120}},\ \bibinfo {pages} {2296} (\bibinfo {year} {2004})}\BibitemShut
  {NoStop}%
\bibitem [{\citenamefont {Tscherbul}\ \emph
  {et~al.}(2009{\natexlab{a}})\citenamefont {Tscherbul}, \citenamefont
  {Groenenboom}, \citenamefont {Krems},\ and\ \citenamefont
  {Dalgarno}}]{Tscherbul:09}%
  \BibitemOpen
  \bibfield  {author} {\bibinfo {author} {\bibfnamefont {T.~V.}\ \bibnamefont
  {Tscherbul}}, \bibinfo {author} {\bibfnamefont {G.~C.}\ \bibnamefont
  {Groenenboom}}, \bibinfo {author} {\bibfnamefont {R.~V.}\ \bibnamefont
  {Krems}},\ and\ \bibinfo {author} {\bibfnamefont {A.}~\bibnamefont
  {Dalgarno}},\ }\bibfield  {title} {\bibinfo {title} {Dynamics of
  {OH$(^2\Pi)$-He} collisions in combined electric and magnetic fields},\
  }\href@noop {} {\bibfield  {journal} {\bibinfo  {journal} {Faraday Discuss.}\
  }\textbf {\bibinfo {volume} {142}},\ \bibinfo {pages} {127} (\bibinfo {year}
  {2009}{\natexlab{a}})}\BibitemShut {NoStop}%
\bibitem [{\citenamefont {Tscherbul}\ \emph
  {et~al.}(2009{\natexlab{b}})\citenamefont {Tscherbul}, \citenamefont
  {Suleimanov}, \citenamefont {Aquilanti},\ and\ \citenamefont
  {Krems}}]{Tscherbul:09c}%
  \BibitemOpen
  \bibfield  {author} {\bibinfo {author} {\bibfnamefont {T.~V.}\ \bibnamefont
  {Tscherbul}}, \bibinfo {author} {\bibfnamefont {Y.~V.}\ \bibnamefont
  {Suleimanov}}, \bibinfo {author} {\bibfnamefont {V.}~\bibnamefont
  {Aquilanti}},\ and\ \bibinfo {author} {\bibfnamefont {R.~V.}\ \bibnamefont
  {Krems}},\ }\bibfield  {title} {\bibinfo {title} {Magnetic field modification
  of ultracold molecule--molecule collisions},\ }\href@noop {} {\bibfield
  {journal} {\bibinfo  {journal} {New J. Phys.}\ }\textbf {\bibinfo {volume}
  {11}},\ \bibinfo {pages} {055021} (\bibinfo {year}
  {2009}{\natexlab{b}})}\BibitemShut {NoStop}%
\bibitem [{\citenamefont {Tscherbul}\ and\ \citenamefont
  {Dalgarno}(2010)}]{Tscherbul:10}%
  \BibitemOpen
  \bibfield  {author} {\bibinfo {author} {\bibfnamefont {T.~V.}\ \bibnamefont
  {Tscherbul}}\ and\ \bibinfo {author} {\bibfnamefont {A.}~\bibnamefont
  {Dalgarno}},\ }\bibfield  {title} {\bibinfo {title} {Quantum theory of
  molecular collisions in a magnetic field: Efficient calculations based on the
  total angular momentum representation},\ }\href
  {https://doi.org/10.1063/1.3503500} {\bibfield  {journal} {\bibinfo
  {journal} {J. Chem. Phys.}\ }\textbf {\bibinfo {volume} {133}},\ \bibinfo
  {pages} {184104} (\bibinfo {year} {2010})}\BibitemShut {NoStop}%
\bibitem [{\citenamefont {Tscherbul}\ \emph {et~al.}(2011)\citenamefont
  {Tscherbul}, \citenamefont {K\l{}os},\ and\ \citenamefont
  {Buchachenko}}]{Tscherbul:11}%
  \BibitemOpen
  \bibfield  {author} {\bibinfo {author} {\bibfnamefont {T.~V.}\ \bibnamefont
  {Tscherbul}}, \bibinfo {author} {\bibfnamefont {J.}~\bibnamefont {K\l{}os}},\
  and\ \bibinfo {author} {\bibfnamefont {A.~A.}\ \bibnamefont {Buchachenko}},\
  }\bibfield  {title} {\bibinfo {title} {Ultracold spin-polarized mixtures of
  {${}^{2}\ensuremath{\Sigma}$} molecules with {$S$}-state atoms: Collisional
  stability and implications for sympathetic cooling},\ }\href
  {https://doi.org/10.1103/PhysRevA.84.040701} {\bibfield  {journal} {\bibinfo
  {journal} {Phys. Rev. A}\ }\textbf {\bibinfo {volume} {84}},\ \bibinfo
  {pages} {040701} (\bibinfo {year} {2011})}\BibitemShut {NoStop}%
\bibitem [{\citenamefont {Suleimanov}\ \emph {et~al.}(2012)\citenamefont
  {Suleimanov}, \citenamefont {Tscherbul},\ and\ \citenamefont
  {Krems}}]{Suleimanov:12}%
  \BibitemOpen
  \bibfield  {author} {\bibinfo {author} {\bibfnamefont {Y.~V.}\ \bibnamefont
  {Suleimanov}}, \bibinfo {author} {\bibfnamefont {T.~V.}\ \bibnamefont
  {Tscherbul}},\ and\ \bibinfo {author} {\bibfnamefont {R.~V.}\ \bibnamefont
  {Krems}},\ }\bibfield  {title} {\bibinfo {title} {Efficient method for
  quantum calculations of molecule-molecule scattering properties in a magnetic
  field},\ }\href {https://doi.org/10.1063/1.4733288} {\bibfield  {journal}
  {\bibinfo  {journal} {J. Chem. Phys.}\ }\textbf {\bibinfo {volume} {137}},\
  \bibinfo {pages} {024103} (\bibinfo {year} {2012})}\BibitemShut {NoStop}%
\bibitem [{\citenamefont {Campbell}\ \emph {et~al.}(2009)\citenamefont
  {Campbell}, \citenamefont {Tscherbul}, \citenamefont {Lu}, \citenamefont
  {Tsikata}, \citenamefont {Krems},\ and\ \citenamefont {Doyle}}]{Campbell:09}%
  \BibitemOpen
  \bibfield  {author} {\bibinfo {author} {\bibfnamefont {W.~C.}\ \bibnamefont
  {Campbell}}, \bibinfo {author} {\bibfnamefont {T.~V.}\ \bibnamefont
  {Tscherbul}}, \bibinfo {author} {\bibfnamefont {H.-I.}\ \bibnamefont {Lu}},
  \bibinfo {author} {\bibfnamefont {E.}~\bibnamefont {Tsikata}}, \bibinfo
  {author} {\bibfnamefont {R.~V.}\ \bibnamefont {Krems}},\ and\ \bibinfo
  {author} {\bibfnamefont {J.~M.}\ \bibnamefont {Doyle}},\ }\bibfield  {title}
  {\bibinfo {title} {Mechanism of collisional spin relaxation in
  $^{3}\ensuremath{\Sigma}$ molecules},\ }\href
  {https://doi.org/10.1103/PhysRevLett.102.013003} {\bibfield  {journal}
  {\bibinfo  {journal} {Phys. Rev. Lett.}\ }\textbf {\bibinfo {volume} {102}},\
  \bibinfo {pages} {013003} (\bibinfo {year} {2009})}\BibitemShut {NoStop}%
\bibitem [{\citenamefont {Hummon}\ \emph {et~al.}(2011)\citenamefont {Hummon},
  \citenamefont {Tscherbul}, \citenamefont {K\l{}os}, \citenamefont {Lu},
  \citenamefont {Tsikata}, \citenamefont {Campbell}, \citenamefont {Dalgarno},\
  and\ \citenamefont {Doyle}}]{Hummon:11}%
  \BibitemOpen
  \bibfield  {author} {\bibinfo {author} {\bibfnamefont {M.~T.}\ \bibnamefont
  {Hummon}}, \bibinfo {author} {\bibfnamefont {T.~V.}\ \bibnamefont
  {Tscherbul}}, \bibinfo {author} {\bibfnamefont {J.}~\bibnamefont {K\l{}os}},
  \bibinfo {author} {\bibfnamefont {H.-I.}\ \bibnamefont {Lu}}, \bibinfo
  {author} {\bibfnamefont {E.}~\bibnamefont {Tsikata}}, \bibinfo {author}
  {\bibfnamefont {W.~C.}\ \bibnamefont {Campbell}}, \bibinfo {author}
  {\bibfnamefont {A.}~\bibnamefont {Dalgarno}},\ and\ \bibinfo {author}
  {\bibfnamefont {J.~M.}\ \bibnamefont {Doyle}},\ }\bibfield  {title} {\bibinfo
  {title} {Cold {N~+~NH} collisions in a magnetic trap},\ }\href
  {https://doi.org/10.1103/PhysRevLett.106.053201} {\bibfield  {journal}
  {\bibinfo  {journal} {Phys. Rev. Lett.}\ }\textbf {\bibinfo {volume} {106}},\
  \bibinfo {pages} {053201} (\bibinfo {year} {2011})}\BibitemShut {NoStop}%
\bibitem [{\citenamefont {Bergeat}\ \emph {et~al.}(2015)\citenamefont
  {Bergeat}, \citenamefont {Onvlee}, \citenamefont {Naulin}, \citenamefont
  {van~der Avoird},\ and\ \citenamefont {Costes}}]{Bergeat:15}%
  \BibitemOpen
  \bibfield  {author} {\bibinfo {author} {\bibfnamefont {A.}~\bibnamefont
  {Bergeat}}, \bibinfo {author} {\bibfnamefont {J.}~\bibnamefont {Onvlee}},
  \bibinfo {author} {\bibfnamefont {C.}~\bibnamefont {Naulin}}, \bibinfo
  {author} {\bibfnamefont {A.}~\bibnamefont {van~der Avoird}},\ and\ \bibinfo
  {author} {\bibfnamefont {M.}~\bibnamefont {Costes}},\ }\bibfield  {title}
  {\bibinfo {title} {Quantum dynamical resonances in low-energy {CO($j =
  0$)~+~He} inelastic collisions},\ }\href {https://doi.org/10.1038/nchem.2204}
  {\bibfield  {journal} {\bibinfo  {journal} {Nat. Chem.}\ }\textbf {\bibinfo
  {volume} {7}},\ \bibinfo {pages} {349} (\bibinfo {year} {2015})}\BibitemShut
  {NoStop}%
\bibitem [{\citenamefont {Vogels}\ \emph {et~al.}(2015)\citenamefont {Vogels},
  \citenamefont {Onvlee}, \citenamefont {Chefdeville}, \citenamefont {van~der
  Avoird}, \citenamefont {Groenenboom},\ and\ \citenamefont {van~de
  Meerakker}}]{Vogels:15}%
  \BibitemOpen
  \bibfield  {author} {\bibinfo {author} {\bibfnamefont {S.~N.}\ \bibnamefont
  {Vogels}}, \bibinfo {author} {\bibfnamefont {J.}~\bibnamefont {Onvlee}},
  \bibinfo {author} {\bibfnamefont {S.}~\bibnamefont {Chefdeville}}, \bibinfo
  {author} {\bibfnamefont {A.}~\bibnamefont {van~der Avoird}}, \bibinfo
  {author} {\bibfnamefont {G.~C.}\ \bibnamefont {Groenenboom}},\ and\ \bibinfo
  {author} {\bibfnamefont {S.~Y.~T.}\ \bibnamefont {van~de Meerakker}},\
  }\bibfield  {title} {\bibinfo {title} {Imaging resonances in low-energy
  {NO-He} inelastic collisions},\ }\href@noop {} {\bibfield  {journal}
  {\bibinfo  {journal} {Science}\ }\textbf {\bibinfo {volume} {350}},\ \bibinfo
  {pages} {787} (\bibinfo {year} {2015})}\BibitemShut {NoStop}%
\bibitem [{\citenamefont {Bergeat}\ \emph {et~al.}(2018)\citenamefont
  {Bergeat}, \citenamefont {Chefdeville}, \citenamefont {Costes}, \citenamefont
  {Morales}, \citenamefont {Naulin}, \citenamefont {Even}, \citenamefont
  {K{\l}os},\ and\ \citenamefont {Lique}}]{Bergeat:18}%
  \BibitemOpen
  \bibfield  {author} {\bibinfo {author} {\bibfnamefont {A.}~\bibnamefont
  {Bergeat}}, \bibinfo {author} {\bibfnamefont {S.}~\bibnamefont
  {Chefdeville}}, \bibinfo {author} {\bibfnamefont {M.}~\bibnamefont {Costes}},
  \bibinfo {author} {\bibfnamefont {S.~B.}\ \bibnamefont {Morales}}, \bibinfo
  {author} {\bibfnamefont {C.}~\bibnamefont {Naulin}}, \bibinfo {author}
  {\bibfnamefont {U.}~\bibnamefont {Even}}, \bibinfo {author} {\bibfnamefont
  {J.}~\bibnamefont {K{\l}os}},\ and\ \bibinfo {author} {\bibfnamefont
  {F.}~\bibnamefont {Lique}},\ }\bibfield  {title} {\bibinfo {title}
  {Understanding the quantum nature of low-energy {C$(^3P_j)$~+~He} inelastic
  collisions},\ }\href {https://doi.org/10.1038/s41557-018-0030-y} {\bibfield
  {journal} {\bibinfo  {journal} {Nat. Chem.}\ }\textbf {\bibinfo {volume}
  {10}},\ \bibinfo {pages} {519} (\bibinfo {year} {2018})}\BibitemShut
  {NoStop}%
\bibitem [{\citenamefont {Klein}\ \emph {et~al.}(2017)\citenamefont {Klein},
  \citenamefont {Shagam}, \citenamefont {Skomorowski}, \citenamefont
  {{\.Z}uchowski}, \citenamefont {Pawlak}, \citenamefont {Janssen},
  \citenamefont {Moiseyev}, \citenamefont {van~de Meerakker}, \citenamefont
  {van~der Avoird}, \citenamefont {Koch},\ and\ \citenamefont
  {Narevicius}}]{Klein:17}%
  \BibitemOpen
  \bibfield  {author} {\bibinfo {author} {\bibfnamefont {A.}~\bibnamefont
  {Klein}}, \bibinfo {author} {\bibfnamefont {Y.}~\bibnamefont {Shagam}},
  \bibinfo {author} {\bibfnamefont {W.}~\bibnamefont {Skomorowski}}, \bibinfo
  {author} {\bibfnamefont {P.~S.}\ \bibnamefont {{\.Z}uchowski}}, \bibinfo
  {author} {\bibfnamefont {M.}~\bibnamefont {Pawlak}}, \bibinfo {author}
  {\bibfnamefont {L.~M.~C.}\ \bibnamefont {Janssen}}, \bibinfo {author}
  {\bibfnamefont {N.}~\bibnamefont {Moiseyev}}, \bibinfo {author}
  {\bibfnamefont {S.~Y.~T.}\ \bibnamefont {van~de Meerakker}}, \bibinfo
  {author} {\bibfnamefont {A.}~\bibnamefont {van~der Avoird}}, \bibinfo
  {author} {\bibfnamefont {C.~P.}\ \bibnamefont {Koch}},\ and\ \bibinfo
  {author} {\bibfnamefont {E.}~\bibnamefont {Narevicius}},\ }\bibfield  {title}
  {\bibinfo {title} {Directly probing anisotropy in atom--molecule collisions
  through quantum scattering resonances},\ }\href
  {https://doi.org/10.1038/nphys3904} {\bibfield  {journal} {\bibinfo
  {journal} {Nat. Phys.}\ }\textbf {\bibinfo {volume} {13}},\ \bibinfo {pages}
  {35} (\bibinfo {year} {2017})}\BibitemShut {NoStop}%
\bibitem [{\citenamefont {Vogels}\ \emph {et~al.}(2018)\citenamefont {Vogels},
  \citenamefont {Karman}, \citenamefont {K{\l}os}, \citenamefont {Besemer},
  \citenamefont {Onvlee}, \citenamefont {van~der Avoird}, \citenamefont
  {Groenenboom},\ and\ \citenamefont {van~de Meerakker}}]{Vogels:18}%
  \BibitemOpen
  \bibfield  {author} {\bibinfo {author} {\bibfnamefont {S.~N.}\ \bibnamefont
  {Vogels}}, \bibinfo {author} {\bibfnamefont {T.}~\bibnamefont {Karman}},
  \bibinfo {author} {\bibfnamefont {J.}~\bibnamefont {K{\l}os}}, \bibinfo
  {author} {\bibfnamefont {M.}~\bibnamefont {Besemer}}, \bibinfo {author}
  {\bibfnamefont {J.}~\bibnamefont {Onvlee}}, \bibinfo {author} {\bibfnamefont
  {A.}~\bibnamefont {van~der Avoird}}, \bibinfo {author} {\bibfnamefont
  {G.~C.}\ \bibnamefont {Groenenboom}},\ and\ \bibinfo {author} {\bibfnamefont
  {S.~Y.~T.}\ \bibnamefont {van~de Meerakker}},\ }\bibfield  {title} {\bibinfo
  {title} {Scattering resonances in bimolecular collisions between {NO}
  radicals and {H$_2$} challenge the theoretical gold standard},\ }\href
  {https://doi.org/10.1038/s41557-018-0001-3} {\bibfield  {journal} {\bibinfo
  {journal} {Nat. Chem.}\ }\textbf {\bibinfo {volume} {10}},\ \bibinfo {pages}
  {435} (\bibinfo {year} {2018})}\BibitemShut {NoStop}%
\bibitem [{\citenamefont {Morita}\ \emph {et~al.}(2018)\citenamefont {Morita},
  \citenamefont {Kosicki}, \citenamefont {\ifmmode~\dot{Z}\else
  \.{Z}\fi{}uchowski},\ and\ \citenamefont {Tscherbul}}]{Morita:18}%
  \BibitemOpen
  \bibfield  {author} {\bibinfo {author} {\bibfnamefont {M.}~\bibnamefont
  {Morita}}, \bibinfo {author} {\bibfnamefont {M.~B.}\ \bibnamefont {Kosicki}},
  \bibinfo {author} {\bibfnamefont {P.~S.}\ \bibnamefont {\ifmmode~\dot{Z}\else
  \.{Z}\fi{}uchowski}},\ and\ \bibinfo {author} {\bibfnamefont {T.~V.}\
  \bibnamefont {Tscherbul}},\ }\bibfield  {title} {\bibinfo {title}
  {Atom-molecule collisions, spin relaxation, and sympathetic cooling in an
  ultracold spin-polarized
  {$\mathrm{Rb}(^{2}S)\ensuremath{-}\mathrm{SrF}(^{2}\mathrm{\ensuremath{\Sigma}}^{+})$}
  mixture},\ }\href {https://doi.org/10.1103/PhysRevA.98.042702} {\bibfield
  {journal} {\bibinfo  {journal} {Phys. Rev. A}\ }\textbf {\bibinfo {volume}
  {98}},\ \bibinfo {pages} {042702} (\bibinfo {year} {2018})}\BibitemShut
  {NoStop}%
\bibitem [{\citenamefont {Morita}\ \emph {et~al.}(2020)\citenamefont {Morita},
  \citenamefont {K\l{}os},\ and\ \citenamefont {Tscherbul}}]{Morita:20}%
  \BibitemOpen
  \bibfield  {author} {\bibinfo {author} {\bibfnamefont {M.}~\bibnamefont
  {Morita}}, \bibinfo {author} {\bibfnamefont {J.}~\bibnamefont {K\l{}os}},\
  and\ \bibinfo {author} {\bibfnamefont {T.~V.}\ \bibnamefont {Tscherbul}},\
  }\bibfield  {title} {\bibinfo {title} {Full-dimensional quantum scattering
  calculations on ultracold atom-molecule collisions in magnetic fields: The
  role of molecular vibrations},\ }\href
  {https://doi.org/10.1103/PhysRevResearch.2.043294} {\bibfield  {journal}
  {\bibinfo  {journal} {Phys. Rev. Research}\ }\textbf {\bibinfo {volume}
  {2}},\ \bibinfo {pages} {043294} (\bibinfo {year} {2020})}\BibitemShut
  {NoStop}%
\bibitem [{\citenamefont {Lara}\ \emph {et~al.}(2006)\citenamefont {Lara},
  \citenamefont {Bohn}, \citenamefont {Potter}, \citenamefont {Sold\'an},\ and\
  \citenamefont {Hutson}}]{Lara:06}%
  \BibitemOpen
  \bibfield  {author} {\bibinfo {author} {\bibfnamefont {M.}~\bibnamefont
  {Lara}}, \bibinfo {author} {\bibfnamefont {J.~L.}\ \bibnamefont {Bohn}},
  \bibinfo {author} {\bibfnamefont {D.}~\bibnamefont {Potter}}, \bibinfo
  {author} {\bibfnamefont {P.}~\bibnamefont {Sold\'an}},\ and\ \bibinfo
  {author} {\bibfnamefont {J.~M.}\ \bibnamefont {Hutson}},\ }\bibfield  {title}
  {\bibinfo {title} {Ultracold {Rb-OH} collisions and prospects for sympathetic
  cooling},\ }\href {https://doi.org/10.1103/PhysRevLett.97.183201} {\bibfield
  {journal} {\bibinfo  {journal} {Phys. Rev. Lett.}\ }\textbf {\bibinfo
  {volume} {97}},\ \bibinfo {pages} {183201} (\bibinfo {year}
  {2006})}\BibitemShut {NoStop}%
\bibitem [{\citenamefont {Wallis}\ and\ \citenamefont
  {Hutson}(2009)}]{Wallis:09}%
  \BibitemOpen
  \bibfield  {author} {\bibinfo {author} {\bibfnamefont {A.~O.~G.}\
  \bibnamefont {Wallis}}\ and\ \bibinfo {author} {\bibfnamefont {J.~M.}\
  \bibnamefont {Hutson}},\ }\bibfield  {title} {\bibinfo {title} {Production of
  ultracold {NH} molecules by sympathetic cooling with {Mg}},\ }\href
  {https://doi.org/10.1103/PhysRevLett.103.183201} {\bibfield  {journal}
  {\bibinfo  {journal} {Phys. Rev. Lett.}\ }\textbf {\bibinfo {volume} {103}},\
  \bibinfo {pages} {183201} (\bibinfo {year} {2009})}\BibitemShut {NoStop}%
\bibitem [{\citenamefont {Pavlovic}\ \emph {et~al.}(2009)\citenamefont
  {Pavlovic}, \citenamefont {Tscherbul}, \citenamefont {Sadeghpour},
  \citenamefont {Groenenboom},\ and\ \citenamefont {Dalgarno}}]{Pavlovic:09}%
  \BibitemOpen
  \bibfield  {author} {\bibinfo {author} {\bibfnamefont {Z.}~\bibnamefont
  {Pavlovic}}, \bibinfo {author} {\bibfnamefont {T.~V.}\ \bibnamefont
  {Tscherbul}}, \bibinfo {author} {\bibfnamefont {H.~R.}\ \bibnamefont
  {Sadeghpour}}, \bibinfo {author} {\bibfnamefont {G.~C.}\ \bibnamefont
  {Groenenboom}},\ and\ \bibinfo {author} {\bibfnamefont {A.}~\bibnamefont
  {Dalgarno}},\ }\bibfield  {title} {\bibinfo {title} {Cold collisions of
  {OH($^2\Pi$)} molecules with {He} atoms in external fields},\ }\href
  {https://doi.org/10.1021/jp904512r} {\bibfield  {journal} {\bibinfo
  {journal} {J. Phys. Chem. A}\ }\textbf {\bibinfo {volume} {113}},\ \bibinfo
  {pages} {14670} (\bibinfo {year} {2009})}\BibitemShut {NoStop}%
\bibitem [{\citenamefont {{\.Z}uchowski}\ and\ \citenamefont
  {Hutson}(2011)}]{Zuchowski:11}%
  \BibitemOpen
  \bibfield  {author} {\bibinfo {author} {\bibfnamefont {P.~S.}\ \bibnamefont
  {{\.Z}uchowski}}\ and\ \bibinfo {author} {\bibfnamefont {J.~M.}\ \bibnamefont
  {Hutson}},\ }\bibfield  {title} {\bibinfo {title} {Cold collisions of
  {N($^4$S)} atoms and {NH($^3\Sigma$)} molecules in magnetic fields},\ }\href
  {https://doi.org/10.1039/C0CP01447H} {\bibfield  {journal} {\bibinfo
  {journal} {Phys. Chem. Chem. Phys.}\ }\textbf {\bibinfo {volume} {13}},\
  \bibinfo {pages} {3669} (\bibinfo {year} {2011})}\BibitemShut {NoStop}%
\bibitem [{\citenamefont {Tscherbul}\ and\ \citenamefont
  {Krems}(2006{\natexlab{a}})}]{Tscherbul:06}%
  \BibitemOpen
  \bibfield  {author} {\bibinfo {author} {\bibfnamefont {T.~V.}\ \bibnamefont
  {Tscherbul}}\ and\ \bibinfo {author} {\bibfnamefont {R.~V.}\ \bibnamefont
  {Krems}},\ }\bibfield  {title} {\bibinfo {title} {Controlling electronic spin
  relaxation of cold molecules with electric fields},\ }\href
  {https://doi.org/10.1103/PhysRevLett.97.083201} {\bibfield  {journal}
  {\bibinfo  {journal} {Phys. Rev. Lett.}\ }\textbf {\bibinfo {volume} {97}},\
  \bibinfo {pages} {083201} (\bibinfo {year} {2006}{\natexlab{a}})}\BibitemShut
  {NoStop}%
\bibitem [{\citenamefont {Morita}\ \emph {et~al.}(2017)\citenamefont {Morita},
  \citenamefont {K\l{}os}, \citenamefont {Buchachenko},\ and\ \citenamefont
  {Tscherbul}}]{Morita:17}%
  \BibitemOpen
  \bibfield  {author} {\bibinfo {author} {\bibfnamefont {M.}~\bibnamefont
  {Morita}}, \bibinfo {author} {\bibfnamefont {J.}~\bibnamefont {K\l{}os}},
  \bibinfo {author} {\bibfnamefont {A.~A.}\ \bibnamefont {Buchachenko}},\ and\
  \bibinfo {author} {\bibfnamefont {T.~V.}\ \bibnamefont {Tscherbul}},\
  }\bibfield  {title} {\bibinfo {title} {Cold collisions of heavy
  $^{2}\mathrm{\ensuremath{\Sigma}}$ molecules with alkali-metal atoms in a
  magnetic field: Ab initio analysis and prospects for sympathetic cooling of
  {SrOH}$(^{2}{\mathrm{\ensuremath{\Sigma}}}^{+})$ by {Li$(^{2}S)$}},\ }\href
  {https://doi.org/10.1103/PhysRevA.95.063421} {\bibfield  {journal} {\bibinfo
  {journal} {Phys. Rev. A}\ }\textbf {\bibinfo {volume} {95}},\ \bibinfo
  {pages} {063421} (\bibinfo {year} {2017})}\BibitemShut {NoStop}%
\bibitem [{\citenamefont {Dulieu}\ and\ \citenamefont
  {Osterwalder}(2018)}]{ColdChem:18}%
  \BibitemOpen
  \bibinfo {editor} {\bibfnamefont {O.}~\bibnamefont {Dulieu}}\ and\ \bibinfo
  {editor} {\bibfnamefont {A.}~\bibnamefont {Osterwalder}},\ eds.,\ \bibinfo
  {title} {{\it {C}old {C}hemistry: {M}olecular {S}cattering and {R}eactivity
  {N}ear {A}bsolute {Z}ero}}\ (\bibinfo  {publisher} {Royal Society of
  Chemistry},\ \bibinfo {year} {2018})\BibitemShut {NoStop}%
\bibitem [{\citenamefont {Tscherbul}(2012)}]{Tscherbul:12}%
  \BibitemOpen
  \bibfield  {author} {\bibinfo {author} {\bibfnamefont {T.~V.}\ \bibnamefont
  {Tscherbul}},\ }\bibfield  {title} {\bibinfo {title} {Total-angular-momentum
  representation for atom-molecule collisions in electric fields},\ }\href
  {https://doi.org/10.1103/PhysRevA.85.052710} {\bibfield  {journal} {\bibinfo
  {journal} {Phys. Rev. A}\ }\textbf {\bibinfo {volume} {85}},\ \bibinfo
  {pages} {052710} (\bibinfo {year} {2012})}\BibitemShut {NoStop}%
\bibitem [{\citenamefont {Arthurs}\ and\ \citenamefont
  {Dalgarno}(1960)}]{Arthurs:60}%
  \BibitemOpen
  \bibfield  {author} {\bibinfo {author} {\bibfnamefont {A.~M.}\ \bibnamefont
  {Arthurs}}\ and\ \bibinfo {author} {\bibfnamefont {A.}~\bibnamefont
  {Dalgarno}},\ }\bibfield  {title} {\bibinfo {title} {The theory of scattering
  by a rigid rotator},\ }\href {http://www.jstor.org/stable/2413933} {\bibfield
   {journal} {\bibinfo  {journal} {Proc. R. Soc. London Ser. A}\ }\textbf
  {\bibinfo {volume} {256}},\ \bibinfo {pages} {540} (\bibinfo {year}
  {1960})}\BibitemShut {NoStop}%
\bibitem [{\citenamefont {Tscherbul}\ \emph {et~al.}(2007)\citenamefont
  {Tscherbul}, \citenamefont {K\l{}os}, \citenamefont {Rajchel},\ and\
  \citenamefont {Krems}}]{Tscherbul:07}%
  \BibitemOpen
  \bibfield  {author} {\bibinfo {author} {\bibfnamefont {T.~V.}\ \bibnamefont
  {Tscherbul}}, \bibinfo {author} {\bibfnamefont {J.}~\bibnamefont {K\l{}os}},
  \bibinfo {author} {\bibfnamefont {L.}~\bibnamefont {Rajchel}},\ and\ \bibinfo
  {author} {\bibfnamefont {R.~V.}\ \bibnamefont {Krems}},\ }\bibfield  {title}
  {\bibinfo {title} {Fine and hyperfine interactions in cold {YbF-He}
  collisions in electromagnetic fields},\ }\href
  {https://doi.org/10.1103/PhysRevA.75.033416} {\bibfield  {journal} {\bibinfo
  {journal} {Phys. Rev. A}\ }\textbf {\bibinfo {volume} {75}},\ \bibinfo
  {pages} {033416} (\bibinfo {year} {2007})}\BibitemShut {NoStop}%
\bibitem [{\citenamefont {Wang}\ \emph {et~al.}(2021)\citenamefont {Wang},
  \citenamefont {Frye}, \citenamefont {Su}, \citenamefont {Can}, \citenamefont
  {Liu}, \citenamefont {Zhang}, \citenamefont {Yang}, \citenamefont {Hutson},
  \citenamefont {Zhao}, \citenamefont {Bai},\ and\ \citenamefont
  {Pan}}]{Wang:21}%
  \BibitemOpen
  \bibfield  {author} {\bibinfo {author} {\bibfnamefont {X.-Y.}\ \bibnamefont
  {Wang}}, \bibinfo {author} {\bibfnamefont {M.~D.}\ \bibnamefont {Frye}},
  \bibinfo {author} {\bibfnamefont {Z.}~\bibnamefont {Su}}, \bibinfo {author}
  {\bibfnamefont {J.}~\bibnamefont {Can}}, \bibinfo {author} {\bibfnamefont
  {L.}~\bibnamefont {Liu}}, \bibinfo {author} {\bibfnamefont {D.-C.}\
  \bibnamefont {Zhang}}, \bibinfo {author} {\bibfnamefont {G.}~\bibnamefont
  {Yang}}, \bibinfo {author} {\bibfnamefont {J.~M.}\ \bibnamefont {Hutson}},
  \bibinfo {author} {\bibfnamefont {B.}~\bibnamefont {Zhao}}, \bibinfo {author}
  {\bibfnamefont {C.-L.}\ \bibnamefont {Bai}},\ and\ \bibinfo {author}
  {\bibfnamefont {J.-W.}\ \bibnamefont {Pan}},\ }\bibfield  {title} {\bibinfo
  {title} {Magnetic {Feshbach} resonances in collisions of
  $^{23}${Na}$^{40}${K} with $^{40}${K}},\ }\href@noop {} {\bibfield  {journal}
  {\bibinfo  {journal} {arXiv:2103.07130v1}\ } (\bibinfo {year}
  {2021})}\BibitemShut {NoStop}%
\bibitem [{\citenamefont {Suleimanov}\ and\ \citenamefont
  {Krems}(2011)}]{Suleimanov:11}%
  \BibitemOpen
  \bibfield  {author} {\bibinfo {author} {\bibfnamefont {Y.~V.}\ \bibnamefont
  {Suleimanov}}\ and\ \bibinfo {author} {\bibfnamefont {R.~V.}\ \bibnamefont
  {Krems}},\ }\bibfield  {title} {\bibinfo {title} {Efficient numerical method
  for locating feshbach resonances of ultracold molecules in external fields},\
  }\href {https://doi.org/10.1063/1.3512627} {\bibfield  {journal} {\bibinfo
  {journal} {J. Chem. Phys.}\ }\textbf {\bibinfo {volume} {134}},\ \bibinfo
  {pages} {014101} (\bibinfo {year} {2011})}\BibitemShut {NoStop}%
\bibitem [{\citenamefont {Zare}(1988)}]{Zare:88}%
  \BibitemOpen
  \bibfield  {author} {\bibinfo {author} {\bibfnamefont {R.~N.}\ \bibnamefont
  {Zare}},\ }\href@noop {} {\emph {\bibinfo {title} {Angular Momentum}}}\
  (\bibinfo  {publisher} {Wiley},\ \bibinfo {year} {1988})\BibitemShut
  {NoStop}%
\bibitem [{\citenamefont {Groenenboom}\ and\ \citenamefont
  {Balakrishnan}(2003)}]{Groenenboom:03}%
  \BibitemOpen
  \bibfield  {author} {\bibinfo {author} {\bibfnamefont {G.~C.}\ \bibnamefont
  {Groenenboom}}\ and\ \bibinfo {author} {\bibfnamefont {N.}~\bibnamefont
  {Balakrishnan}},\ }\bibfield  {title} {\bibinfo {title} {The
  {He-CaH($^2\Sigma^+$) interaction. I. Three-dimensional} ab initio potential
  energy surface},\ }\href {https://doi.org/10.1063/1.1562946} {\bibfield
  {journal} {\bibinfo  {journal} {J. Chem. Phys.}\ }\textbf {\bibinfo {volume}
  {118}},\ \bibinfo {pages} {7380} (\bibinfo {year} {2003})}\BibitemShut
  {NoStop}%
\bibitem [{\citenamefont {Krems}\ \emph {et~al.}(2003)\citenamefont {Krems},
  \citenamefont {Dalgarno}, \citenamefont {Balakrishnan},\ and\ \citenamefont
  {Groenenboom}}]{Krems:03}%
  \BibitemOpen
  \bibfield  {author} {\bibinfo {author} {\bibfnamefont {R.~V.}\ \bibnamefont
  {Krems}}, \bibinfo {author} {\bibfnamefont {A.}~\bibnamefont {Dalgarno}},
  \bibinfo {author} {\bibfnamefont {N.}~\bibnamefont {Balakrishnan}},\ and\
  \bibinfo {author} {\bibfnamefont {G.~C.}\ \bibnamefont {Groenenboom}},\
  }\bibfield  {title} {\bibinfo {title} {Spin-flipping transitions in
  ${}^{2}\ensuremath{\Sigma}$ molecules induced by collisions with
  structureless atoms},\ }\href {https://doi.org/10.1103/PhysRevA.67.060703}
  {\bibfield  {journal} {\bibinfo  {journal} {Phys. Rev. A}\ }\textbf {\bibinfo
  {volume} {67}},\ \bibinfo {pages} {060703} (\bibinfo {year}
  {2003})}\BibitemShut {NoStop}%
\bibitem [{\citenamefont {Balakrishnan}\ \emph {et~al.}(2003)\citenamefont
  {Balakrishnan}, \citenamefont {Groenenboom}, \citenamefont {Krems},\ and\
  \citenamefont {Dalgarno}}]{Balakrishnan:03}%
  \BibitemOpen
  \bibfield  {author} {\bibinfo {author} {\bibfnamefont {N.}~\bibnamefont
  {Balakrishnan}}, \bibinfo {author} {\bibfnamefont {G.~C.}\ \bibnamefont
  {Groenenboom}}, \bibinfo {author} {\bibfnamefont {R.~V.}\ \bibnamefont
  {Krems}},\ and\ \bibinfo {author} {\bibfnamefont {A.}~\bibnamefont
  {Dalgarno}},\ }\bibfield  {title} {\bibinfo {title} {The
  {He-CaH($^2\Sigma^+$) interaction. II. Collisions} at cold and ultracold
  temperatures},\ }\href {https://doi.org/10.1063/1.1562947} {\bibfield
  {journal} {\bibinfo  {journal} {J. Chem. Phys.}\ }\textbf {\bibinfo {volume}
  {118}},\ \bibinfo {pages} {7386} (\bibinfo {year} {2003})}\BibitemShut
  {NoStop}%
\bibitem [{\citenamefont {Tscherbul}\ and\ \citenamefont
  {Krems}(2006{\natexlab{b}})}]{Tscherbul:06b}%
  \BibitemOpen
  \bibfield  {author} {\bibinfo {author} {\bibfnamefont {T.~V.}\ \bibnamefont
  {Tscherbul}}\ and\ \bibinfo {author} {\bibfnamefont {R.~V.}\ \bibnamefont
  {Krems}},\ }\bibfield  {title} {\bibinfo {title} {Manipulating spin-dependent
  interactions in rotationally excited cold molecules with electric fields},\
  }\href {https://doi.org/10.1063/1.2374896} {\bibfield  {journal} {\bibinfo
  {journal} {J. Chem. Phys.}\ }\textbf {\bibinfo {volume} {125}},\ \bibinfo
  {pages} {194311} (\bibinfo {year} {2006}{\natexlab{b}})}\BibitemShut
  {NoStop}%
\bibitem [{\citenamefont {Abrahamsson}\ \emph {et~al.}(2007)\citenamefont
  {Abrahamsson}, \citenamefont {Tscherbul},\ and\ \citenamefont
  {Krems}}]{Abrahamsson:07}%
  \BibitemOpen
  \bibfield  {author} {\bibinfo {author} {\bibfnamefont {E.}~\bibnamefont
  {Abrahamsson}}, \bibinfo {author} {\bibfnamefont {T.~V.}\ \bibnamefont
  {Tscherbul}},\ and\ \bibinfo {author} {\bibfnamefont {R.~V.}\ \bibnamefont
  {Krems}},\ }\bibfield  {title} {\bibinfo {title} {Inelastic collisions of
  cold polar molecules in nonparallel electric and magnetic fields},\ }\href
  {https://doi.org/10.1063/1.2748770} {\bibfield  {journal} {\bibinfo
  {journal} {J. Chem. Phys.}\ }\textbf {\bibinfo {volume} {127}},\ \bibinfo
  {pages} {044302} (\bibinfo {year} {2007})}\BibitemShut {NoStop}%
\end{thebibliography}%

\end{document}